\documentclass[preprint,onecolumn,nofootinbib]{revtex4}
\pdfoutput=1
\usepackage{float}
\usepackage[colorlinks=true,linkcolor=blue,urlcolor=blue,filecolor=black,citecolor=red,pdfstartview=FitV,pdftitle={},pdfsubject={},pdfkeywords={},pdfpagemode=None,bookmarksopen=true]{hyperref}
\usepackage{graphicx}
\usepackage{amsmath}
\usepackage{mathrsfs}
\usepackage{amsfonts}
\usepackage{amssymb,ulem}
\usepackage{color,xcolor}%
\usepackage{subfigure}
\usepackage{dcolumn}
\usepackage[T1]{fontenc}
\usepackage{array}
\usepackage{booktabs}

\begin{document}
\title{Collective dynamics in holographic fractonic solids}

\author{Ling-Zheng Xia$^{1}$}
\thanks{xialingzheng@mail.dlut.edu.cn}
\author{Lixin Xu$^{1}$}
\thanks{lxxu@dlut.edu.cn}
\author{Wei-Jia Li$^{1}$}
\thanks{weijiali@dlut.edu.cn (corresponding author)}

\affiliation{$^{1}$Institute of Theoretical Physics, School of Physics, Dalian University of Technology, Dalian 116024, China.
}

\begin{abstract}
Fractonic phases of matter, a class of states in which collective excitations with constrained mobility exist, were originally discovered in the study of quantum error-correcting codes in solvable lattice spin models such as Haah's code and the X-cube model. Recently, they have also drawn the attention of the high-energy physics community due to the UV/IR mixing that arises when coarse-graining these lattice models. In this work, we consider a (3+1)-dimensional holographic model of fractonic solids and investigate the low-energy collective dynamics systematically. By computing the quasinormal modes of black holes, we obtain all the hydrodynamic excitations on the boundary, including two acoustic phonons, a longitudinal diffusive mode, and a subdiffusive collective mode with the dispersion $\omega \sim-ik^4$. In addition, it is found that the latter remains gapless when translational symmetry is explicitly broken. These results suggest that the subdiffusive mode is inherently protected by the crystal-dipole symmetry in solids and is qualitatively unaffected by broken spacetime symmetries.
\end{abstract}

\maketitle
\tableofcontents
\section{Introduction}\label{section1}
Fractons, an exotic class of elementary excitations exhibiting constrained mobility, have attracted broad attention in the study of many-body quantum matter \cite{Chamon:2004lew,Dubovsky:2005xd,Haah:2011drr,Vijay:2015mka,Vijay:2016phm,Qi:2022seq,Jain:2021ibh,Bidussi:2021nmp}.\footnote{The term ``fracton'' was earlier introduced in particle physics to describe a type of fractionally charged colorless bound system \cite{Khlopov:1981wm}. While, in condensed matter physics, it has been introduced to refer to quasiparticles with restricted mobility much later.} The concept of fractons originated in the study of exactly solvable lattice models for quantum error correction, most notably Haah's code \cite{Haah:2011drr} and the X-cube model \cite{Vijay:2016phm}. In recent years, these systems have also garnered significant interest within the high-energy physics community. This interest is largely driven by the phenomenon of UV/IR mixing observed when attempting to describe these lattice models using continuum field theories, where the low-energy physics remains sensitive to microscopic lattice details \cite{Seiberg:2020wsg,You:2021tmm,Gorantla:2021bda}.

Recently, it has been proposed that the constrained motion of collective modes arises because, in addition to the monopole charge, multipole moments such as the dipole moment are also conserved \cite{Pretko:2018jbi,Pretko:2020cko}, making individual charge carriers immobile while allowing certain neutral bound states to spread. This mechanism gives rise to a family of subdiffusive modes. Such a dramatic slowing of collective dynamics is a manifestation of ergodicity breaking in the system \cite{Prem:2017qcp,Pai:2018gyd}. Furthermore, the inherent stability of these excitations also implies potential applications in quantum information storage and processing \cite{Brown:2019hxw,Terhal:2013vbm,Bravyi:2011faf}.
Several reviews summarize developments in these areas (see \cite{Gromov:2022cxa,Nandkishore:2018sel,Pretko:2020cko}).

The effort to formulate a universal description of fractonic phases has led to the development of the so-called fracton hydrodynamics. This framework has been built progressively, starting from simple systems with only charge and dipole conservation \cite{Gromov:2020yoc}, and later adding other key ingredients like momentum \cite{Grosvenor:2021rrt,Glorioso:2021bif,Burchards:2022lqr,Guo:2022ixk,Hart_2022,Glorioso:2023chm,Osborne:2021mej} as well as energy \cite{Glodkowski:2022xje,Jain:2023nbf,Jain:2024kri,Jain:2024ngx}. 

The dipole symmetry of fractons is related to the breaking of continuous spatial translations \cite{Pretko:2020cko,Glorioso:2023chm,Gromov:2018nbv}. Since this symmetry breaking is a ubiquitous feature of solid systems, occurring either spontaneously or explicitly through mechanisms such as lattices, impurities, or defects \cite{Lubensky,RevModPhys.60.1129}, solids serve as a crucial platform for the realization of fractons. However, embedding fractons within the solid-state context raises a natural question: how do the Goldstone modes of a solid (acoustic phonons) coexist and interact with the subdiffusive mode of a fractonic system? The study of fractonic solids will enrich our understanding of fractonic phases as well as the interplay between the dipole symmetry and spacetime symmetries.

During the past few decades, the AdS/CFT correspondence, or holographic duality, has become a powerful tool for tackling real-time dynamics of strongly correlated many-body systems \cite{Ammon_Erdmenger_2015,Baggioli:2019rr,Maldacena:1997re,Witten:1998qj}. This approach has proven to be particularly powerful in condensed matter physics, especially for understanding exotic phases of matter where traditional perturbative methods fail \cite{Hartnoll:2016apf,Zaanen_Liu_Sun_Schalm_2015}. A key point for modeling properties of solids is to break the translational symmetry spontaneously. Among the various holographic systems implementing this symmetry breaking pattern, the holographic axion model stands out as perhaps the most economical. For a detailed review of this model, one can refer to \cite{Baggioli:2021xuv}.

Inspired by recent theoretical developments, especially the hydrodynamic studies of collective modes in \cite{Jain:2024ngx,Glodkowski:2025tnv,Stahl:2023prt} and the holographic subdiffusion model in \cite{Ganesan:2020wvm}, we investigate the collective dynamics of fractonic solids through a generalized holographic axion model coupled with monopole and crystal-dipole gauge fields in the bulk. In particular, we focus on the intricate interplay among the subdiffusive collective mode, energy-momentum fluctuations, and acoustic phonons in solid systems.

To begin with, we review the effective theory formulation of fractonic solids. In the framework of effective field theory (EFT), the low-energy dynamics of a solid can be described by introducing a set of scalar fields $\{\phi^I,\,I=1,2,\dots,d\}$, which serve as co-moving coordinates for the constituent elements of the solid. In the equilibrium state, their vacuum expectation values are linearly dependent on the spatial coordinates \cite{Nicolis:2013lma},
\begin{equation}
    \langle \phi^I \rangle=\delta^I_ix^i, \label{vev}
\end{equation}
where lowercase Latin indices denote spatial directions and uppercase indices label the internal field space. Although this configuration spontaneously breaks the translational invariance, homogeneity is preserved by imposing an internal shift symmetry
\begin{equation}
    \phi^I \to \phi^I+c^I.
\end{equation} 
Then, one can construct the general low energy effective action that is invariant under both the spatial translations and  internal shifts. The vacuum expectation value \eqref{vev} spontaneously breaks the symmetry group down to the diagonal subgroup. The fluctuations of the scalar fields around their equilibrium positions $\delta\phi^I=\phi^I-\langle \phi^I \rangle$ are precisely the Goldstone modes of broken translations. In this way, the phonon dynamics and the associated elasticity of solids can be reproduced in this formulation \cite{Nicolis:2013lma,Nicolis:2015sra,Esposito:2017qpj}. 

In addition, a complex scalar $\Psi$ is introduced to describe the fractonic matter, associated with the monopole and crystal-dipole symmetry \cite{Jain:2024ngx,Glodkowski:2025tnv}
\begin{equation}
    \Psi \to e^{i(a+b^I\phi^I)}\Psi,
\end{equation}
where $a$ and $b^I$ are constant parameters corresponding to the monopole and crystal-dipole transformations, respectively. To derive the Ward identities of the conserved quantities, one can introduce two gauge fields, $\mathbb{A}_\mu$ and $\mathbb{A}^I_\mu$, which allow us to probe the response of the system and identify the conserved currents. Their gauge transformations are imposed as
 \begin{equation}
     \begin{cases}
         \mathbb{A}_\mu \to \mathbb{A}_\mu+\partial_\mu a(x),\\
         \mathbb{A}^I_\mu \to \mathbb{A}^I_\mu+\partial_\mu b^I(x),
     \end{cases}
 \end{equation}
where $a(x)$ and $b^I(x)$  are arbitrary local functions. Then, introducing the covariant derivative $D_\mu\equiv\partial_\mu- i\tilde{\mathbb{A}}_\mu=\partial_\mu-i(\mathbb{A}_\mu+\phi^I\mathbb{A}^I_\mu)$ together with the spatial covariant operator
\begin{equation}
    D^{IJ}(\Psi,\Psi)\equiv \Psi D^{(I}D^{J)} \Psi-D^I\Psi D^J\Psi-i\partial^\mu\phi^{(I}A^{J)}_\mu \Psi^2,\quad\text{with}\ D^I\equiv\partial^\mu\phi^ID_\mu,
\end{equation}
one can construct a crystal-dipole invariant Lagrangian \cite{Jain:2024ngx}.
Moreover, varying the effective action of the theory with respect to the metric and the two gauge fields yields the expression \cite{Glodkowski:2025tnv}
\begin{equation}
    \delta \mathbb{S}=\int \mathrm{d}^{d+1}x \bigg(\frac{1}{2}T^{\mu\nu}\delta g_{\mu\nu}+ J^\mu\delta \tilde{\mathbb{A}}_\mu+ J^{I\mu}\delta \mathbb{A}_\mu^I\bigg).
\end{equation}
The total variations above are given by
\begin{align}
    \delta g_{\mu\nu}=&\mathcal{L}_\xi g_{\mu\nu},\\
    \delta \tilde{\mathbb{A}}_\mu=&\mathcal{L}_\xi \tilde{\mathbb{A}}_\mu+\partial_\mu a(x)+\phi^I\partial_\mu b^I(x)\\
    \delta \mathbb{A}^I_\mu=&\mathcal{L}_\xi \mathbb{A}^I_\mu+\partial_\mu b^I(x),
\end{align}
where $\mathcal{L}_\xi$ denotes the Lie derivative with respect to a vector field $\xi^\mu$. The corresponding Ward identities related to spacetime diffeomorphism, monopole transformation, and crystal-dipole transformation are expressed as
\begin{align}
     \partial_\mu T^{\mu\nu}&=\mathscr{F}^\nu_{\ \mu}  J^\mu+\mathbb{F}^{I \nu}_{\ \ \, \mu} J^{I \mu}, \label{eqdiffeo}\\
      \partial_\mu J^\mu&=0, \label{eqmonopole}\\
    \partial_\mu J^{I \mu}&=-J^\mu\partial_\mu\phi^I, \label{eqdipole}
\end{align}
where 
\begin{align} \mathscr{F}_{\mu\nu}&\equiv \tilde{\mathbb{F}}_{\mu\nu}+2\mathbb{A}^I_{[\mu}\partial_{\nu]}\phi^I=2\partial_{[\mu}\tilde{\mathbb{A}}_{\nu]}+2\mathbb{A}^I_{[\mu}\partial_{\nu]}\phi^I, \label{crystaldipole}\\
\mathbb{F}^I_{\mu\nu}&\equiv2\partial_{[\mu}\mathbb{A}^I_{\nu]}. \label{dipole}
\end{align}
It is worth pointing out that the effective field strength $\mathscr{F}_{\mu\nu}$ is not the standard electromagnetic tensor. Instead, it includes contributions from the crystal-dipole gauge field and the solid structure.

The rest of this paper will be structured as follows: In Section \ref{section2}, we introduce the holographic model of fractonic solids inspired by the above discussion and obtain the black hole solution. In Section \ref{section3}, the dispersion relations of collective modes are obtained by computing the low-lying quasinormal modes of the black hole by breaking spatial translational symmetry spontaneously and explicitly. Finally, we provide a discussion and outlook in Section \ref{section4}.

\section{Holographic model for fractonic solids}\label{section2}

In this section, we construct a (3+1)-dimensional holographic model of fractonic solids guided by the framework outlined in the previous section. The holographic dictionary states that currents in the boundary theory should be sourced by the corresponding fields in the bulk. Therefore, one can impose the following map from the boundary to the bulk:
\begin{equation}
\tilde{\mathbb{A}}_\mu\mapsto A_a, \quad \mathbb{A}^I_\mu\mapsto  A_a^I,
\end{equation}
where the subscript `$a$' denotes the spacetime index in the bulk. To break the translational symmetry,
we further introduce a set of massless bulk scalars $\{\Phi^I,\,I=1,2\}$ (which are also called `axion' fields in much of the literature) possessing an internal shift symmetry $\Phi^I\to \Phi^I+c^I$.  
  Then, the bulk action can be written as \cite{Jain:2024ngx}
\begin{equation}
    S = \int \mathrm{d}^4x \sqrt{-g} \bigg[R-2\Lambda- 2m^2 V(X)-\frac{1}{2}\mathcal{F}_{ab}\mathcal{F}^{ab}-\frac{1}{2} F^I_{ab}F^{Iab}
     \bigg],
\end{equation}
where $R$ is the Ricci scalar, $\Lambda$ is the negative cosmological constant, $m^2$ is an effective coupling with the dimension of mass squared, and $V(X)$ is a general function of $X\equiv \dfrac{1}{2}\nabla_a\Phi^I\nabla^a\Phi^I$. Following \eqref{crystaldipole} and \eqref{dipole}, the field strengths $\mathcal{F}_{ab}$ and $F^I_{ab}$ in the bulk take the forms
\begin{equation}
 \mathcal{F}_{ab}=2\nabla_{[a} A_{b]}+2A^I_{[a}\nabla_{b]}\Phi^I,\quad F^I_{ab}=2\nabla_{[a} A^I_{b]}.
\end{equation}
They provide the external sources of the currents $J^\mu$ and $J^{I\mu}$ on the boundary. Under variation of the bulk fields, equations of motion (EOMs) are given by
\begin{align}
 0=&\ G_{ab}+\Lambda g_{ab}-  F^I_{c a}F^{Ic}_{b}+ \frac{1}{4}g_{ab}F^I_{dc}F^{Idc}-  \mathcal{F}_{c a}\mathcal{F}^{c}_{b}+ \frac{1}{4}g_{ab}\mathcal{F}_{dc}\mathcal{F}^{dc}\nonumber\\&+m^2 g_{ab}  V(X)-  m^2V'(X)\nabla_a\Phi^I\nabla_b\Phi^I,\\
0=&\ m^2\nabla_a(V'(X)\nabla^a\Phi^I)-\frac{1}{2}F^{Iab}\mathcal{F}_{ab},\\
0=&\ \nabla_a \mathcal{F}^{ab}, \label{eqA}\\
 0=&\ \nabla_a F^{Iab}+\nabla_a\Phi^I\mathcal{F}^{ab}.\label{eqAI}
\end{align} 

A specific realization of translational symmetry breaking that retains the system's homogeneity and isotropy is imposed by the following background profile for the scalars \cite{Vegh:2013sk,Andrade:2013gsa}
\begin{align}
\bar{\Phi}^I=\delta_i^Ix^i.\label{scalarpro}
\end{align}
In the ingoing Eddington–Finkelstein (EF) coordinates, the ansatz on the black hole solution can be taken as
\begin{equation}
    \mathrm{d}s^2=\dfrac{1}{u^2}[-f(u) dt^2-2dtdu+dx^2+dy^2],
\end{equation}
where $u\in[0,u_h]$ is the radial coordinate. The AdS boundary is located at $u=0$, and the black hole horizon is at $u=u_h$, defined by the condition $f(u_h)=0$. For a fractonic phase with finite charge density, we can set the background values of the gauge fields,
\begin{align}
   \bar{A}=\bar{A}_t(u) dt, \,\,\text{with} \,\, \bar{A}_t(u) =\mu-\rho u,\quad
   \bar{A}^I=0.
\end{align}
Requiring the temporal component of the gauge field to vanish at the horizon, $\rho=\mu/{u_h}$ should be satisfied. According to the Einstein equation, the emblackening factor is
\begin{equation}
    f(u) =u^3\int^u_{u_h}\mathrm{d}z \bigg[\frac{\Lambda+  m^2V(z^2)}{z^4}+\frac{ \mu^2}{2u_h^2}\bigg].
\end{equation}
Then, the Hawking temperature is read as
\begin{equation}
    T=-\frac{f'(u_h)}{4\pi}=-\frac{2\Lambda+2V(u_h^2)+\rho^2u_h^4}{8\pi u_h},
\end{equation}
which should be viewed as the temperature of the boundary system.

Note that the setup \eqref{scalarpro} cannot ensure that the bulk system is dual to a solid where translations are broken spontaneously. In fact, the way of symmetry breaking very much depends on the choice of $V(X)$. In this work, the benchmark model considered is $V(X)=X^N$. As was firstly realized in \cite{Alberte:2017oqx}, when $N>5/2$, the $x^i$-dependent profile $\bar{\Phi}^I=\delta^I_i\,x^i$ plays the role of the vacuum expectation value of the dual scalar operators in the standard quantization, which exactly coincides with the equilibrium value of the co-moving coordinates $\phi^I$, as defined in Eq.~\eqref{vev}, on the boundary. In contrast, if $N<5/2$, the chosen profile corresponds to external sources that break the symmetry explicitly, hence leading to momentum relaxation \cite{Andrade:2013gsa}.

\section{Hydrodynamic modes}\label{section3}
Having established our holographic model, we now turn to the main objective of this work: studying low-energy collective excitations in a fractonic solid. While the existence of a subdiffusive mode has been anticipated in \cite{Jain:2024ngx}, we provide here explicit calculations to verify its presence and further characterize the other coexisting hydrodynamic excitations in detail. The hydrodynamic modes of the boundary theory are encoded in the quasinormal modes (QNMs) of the black hole, which correspond to the poles of the system's retarded correlators. To extract these modes, we consider small perturbations around the background solution for each of these fields,
\begin{equation*}
    g_{ab}=\bar{g}_{ab}+\delta g_{ab},\quad
\Phi^I=\bar{\Phi}^I+\delta\Phi^I,\quad
    A_a=\bar{A}_a+\delta A_a,\quad
    A^I_a=\bar{A}^I_a+\delta A^I_a,
\end{equation*}
where the fluctuating fields can be expanded as
\begin{equation}
\begin{aligned}
    \delta g_{ab}&=\int \frac{\mathrm{d}^2k\,\mathrm{d}\omega}{(2\pi)^3}\, \dfrac{h_{ab}(\omega,k_i,u)}{u^2}\,e^{-i\omega t+ik_ix^i},\\
    \delta\Phi^I&=\int \frac{\mathrm{d}^2k\,\mathrm{d}\omega}{(2\pi)^3}\,  \delta^I_j\,\Phi_j(\omega,k_i,u)\,e^{-i\omega t+ik_ix^i},\\
    \delta A_a&=\int \frac{\mathrm{d}^2k\,\mathrm{d}\omega}{(2\pi)^3}\,a_a(\omega,k_i,u)\,e^{-i\omega t+ik_ix^i},\\
    \delta A^I_a&=\int \frac{\mathrm{d}^2k\,\mathrm{d}\omega}{(2\pi)^3}\,a^I_{a}(\omega,k_i,u)\,e^{-i\omega t+ik_ix^i}. \label{fourier}
\end{aligned}
\end{equation}
Due to the rotational invariance of the bulk geometry in the $x-y$ plane, we align the wave vector $k$ along the $y-$direction without loss of generality. This choice causes the linearized fluctuations to split into two sectors according to parity under $x\to-x$:
\begin{equation}
\begin{aligned}
\text{Transverse sector (parity odd):}\quad \quad \, \{&\delta g_{xt},\delta g_{xy},\delta g_{xu},\delta\Phi^x,\delta A_{x},\delta A_x^y,\delta A^x_y,\delta A^x_t,\delta A^x_u\},\\
\text{Longitudinal sector (parity even):}\quad \{&\delta g_{uu},\delta g_{tu},\delta g_{yu},\delta g_{tt},\delta g_{ty},\delta g_{xx},\delta g_{yy},\delta\Phi^y,\delta A_{y},\delta A_{t},\delta A_{u},\\& \delta A^y_y,\delta A^y_t,\delta A^y_u,\delta A_x^x\}.
\end{aligned}
\end{equation} 

By solving the coupled linearized EOMs for both sectors (see Appendix \ref{sectionA} for details), subject to the physical boundary conditions of being purely infalling at the horizon and normalizable (sourceless) at the asymptotic boundary, we obtain the QNM spectra of the black hole for finite frequency $\omega$ and momentum $k$. As mentioned in the Introduction, we will realize spontaneous and pseudo-spontaneous breaking scenarios for translations and analyze the low-energy dynamics of fractonic solids. In the numerical computation, we fix $\Lambda=-3$ so that the AdS radius is unity and set the event horizon radius $u_h=1$.

\subsection{Spontaneous breaking of translations}\label{section31}
In this subsection, the case of purely spontaneous symmetry breaking will be considered by imposing $V(X)=X^N\ \text{with}\,\,N>5/2$. Specifically, we fix $N=5$ and show the numerical results.

\begin{figure}[htbp]
    \centering
\includegraphics[width=0.48
    \linewidth]{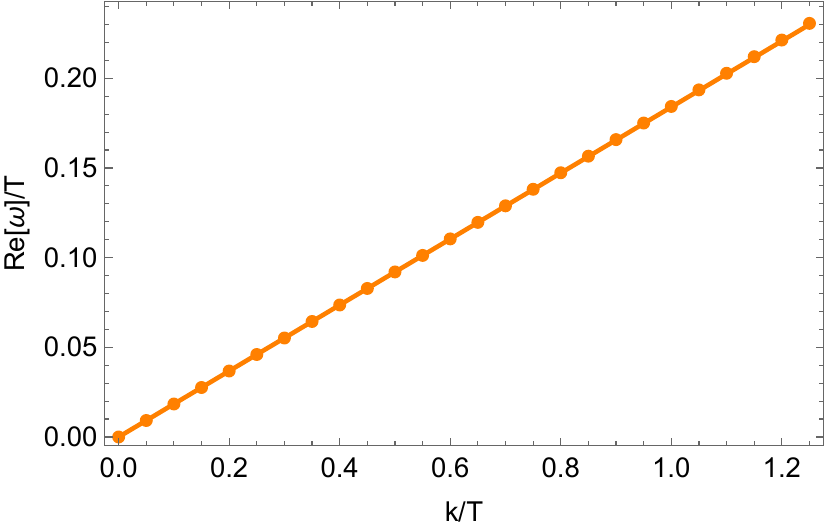}
\includegraphics[width=0.48
    \linewidth]{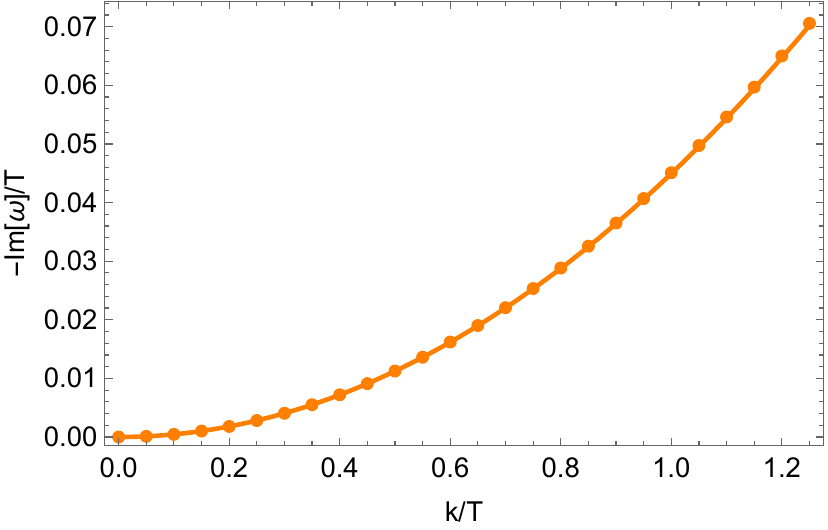}
    \caption{Dispersion relations of transverse hydrodynamic modes, where the dots are the numerical data points and the solid line represents the hydrodynamic prediction \eqref{trsound}. \textbf{Left:} Real part. \textbf{Right:} Imaginary part. Here, we have fixed $\mu/T=1,\ m/T=1.2$. We only plot the branch with positive real part of frequency ($\text{Re}[\omega]>0$). The corresponding negative branch ($\text{Re}[\omega]<0$) is symmetric and has been omitted for clarity. This omission applies to all figures in the following.} 
    \label{transback}
\end{figure}

\textbf{Transverse sector}. FIG.~\ref{transback} shows a pair of sound modes (acoustic phonons) whose dispersion relations at small momenta take the form
\begin{equation}
\omega=\pm v_Tk-i\frac{\Gamma_T}{2}k^2+\cdots, \label{trsound}
\end{equation}
where $v_T$ is the propagation speed of the transverse sounds and $\Gamma_T$ is the sound attenuation. These sound modes are produced by the coupling among $\delta \Phi^x$, $\delta g_{xy}$, and $\delta g_{xt}$. Although the crystal-dipole gauge field $\delta A_a^I$ is also coupled to these fields, it does not lead to any new modes or qualitatively alter the existing ones in this sector. 

\begin{figure}[htbp]
    \centering
\includegraphics[width=0.48
    \linewidth]{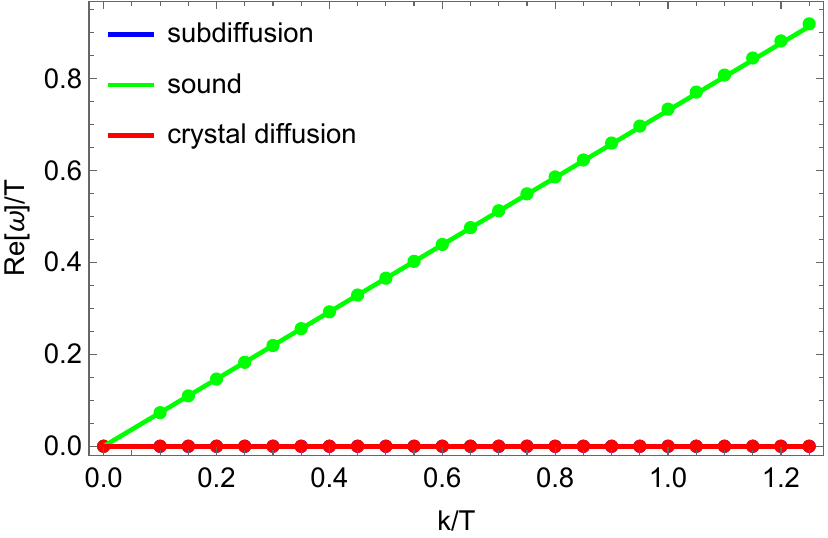}
\includegraphics[width=0.49
    \linewidth]{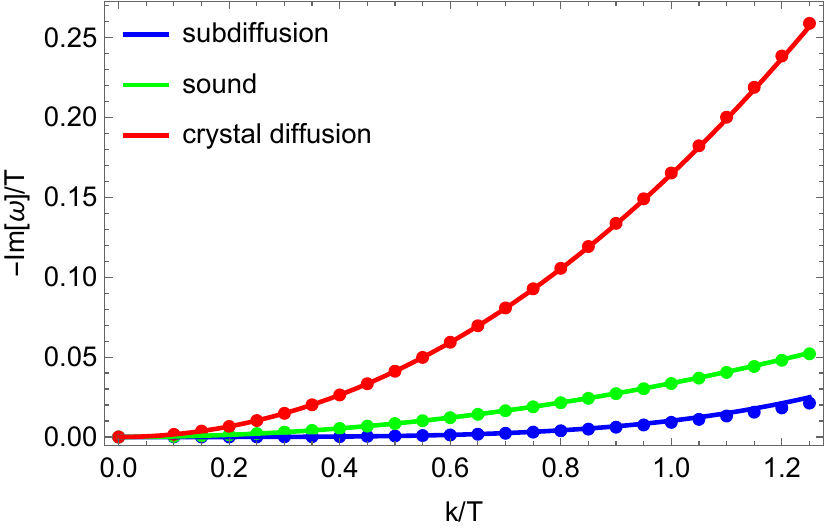}
    \caption{Dispersion relations of longitudinal hydrodynamic modes for $\mu/T=1,\ m/T=1.2$. \textbf{Left:} Real part. \textbf{Right:} Imaginary part. The plotting conventions are the same as in FIG.~\ref{transback}.}
    \label{longiback}
\end{figure}

\textbf{Longitudinal sector}. As shown in FIG.~\ref{longiback}, this sector consists of three types of low-energy modes: a pair of sound modes, a longitudinal diffusive mode\footnote{In some papers (e.g. \cite{Baggioli:2019abx,Baggioli:2020edn}), it is also called crystal diffusion.} and a subdiffusive mode. The dispersion relations of these hydrodynamic modes at small momenta are given by
\begin{equation}
    \begin{cases}
    \text{sound mode}:&\omega=\pm v_Lk-i\dfrac{\Gamma_L}{2}k^2+\cdots,\\
    \text{longitudinal diffusive mode}:&\omega=-iD_L k^2+\cdots, \\
    \text{subdiffusive mode}:&\omega=-i\mathfrak{D}_s k^4+\cdots,\\
    \end{cases}
\end{equation}
where $v_L$ is the propagating speed of the longitudinal phonons, $\Gamma_L$ is the sound attenuation, $D_L$ is the crystal diffusivity and $\mathfrak{D}_s$ is the subdiffusivity. Importantly, the presence of the last mode is the main feature of fractonic phases. This mode arises due to the coupling between $\delta A'_t$ and $\delta A^{y\prime}_t$, where the prime denotes the partial derivative of the radial coordinate $u$. In the charged case, they are coupled to the fluctuations of the metric and axions. However, in the neutral case ($\mu=0$), their EOMs are decoupled from the others, solely giving rise to the subdiffusion (see Appendix \ref{sectionB} for more details).

\begin{figure}[htbp]
    \centering
\includegraphics[width=0.48
    \linewidth]{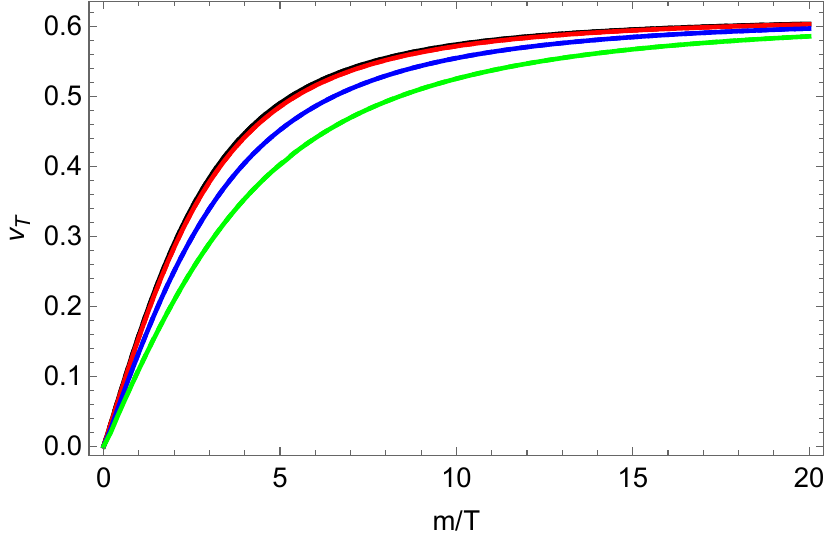}
\includegraphics[width=0.49
    \linewidth]{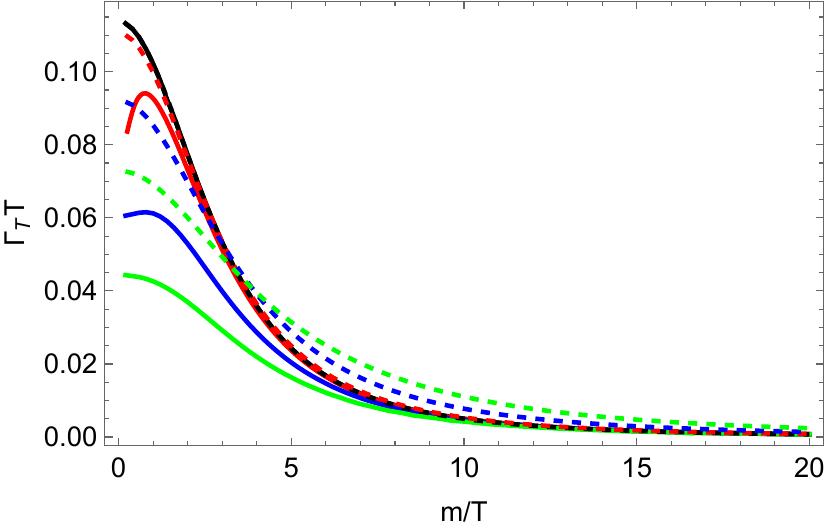}
    \caption{The coefficients in the transverse sound modes as the function of $m/T$ for $\mu/T =\{0, 1, 3, 5\}$ (black, red, blue, green). \textbf{Left:} Transverse sound velocity $v_T$. \textbf{Right:} Dimensionless transverse sound attenuation $\Gamma_TT$. The solid and dashed lines represent the phase with and without $\delta A_a^I$, respectively. The same convention is used for all subsequent plots.}
    \label{transsound}
\end{figure}
\begin{figure}[htbp]
    \centering
\includegraphics[width=0.48
    \linewidth]{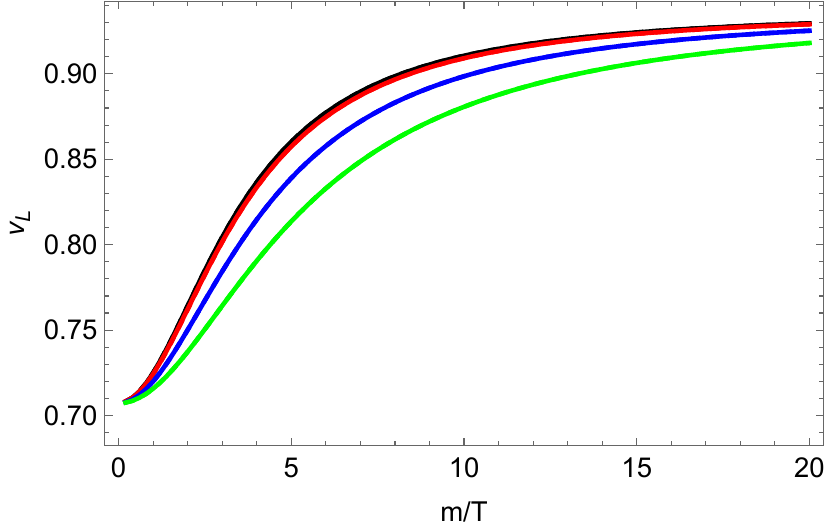}
\includegraphics[width=0.49
    \linewidth]{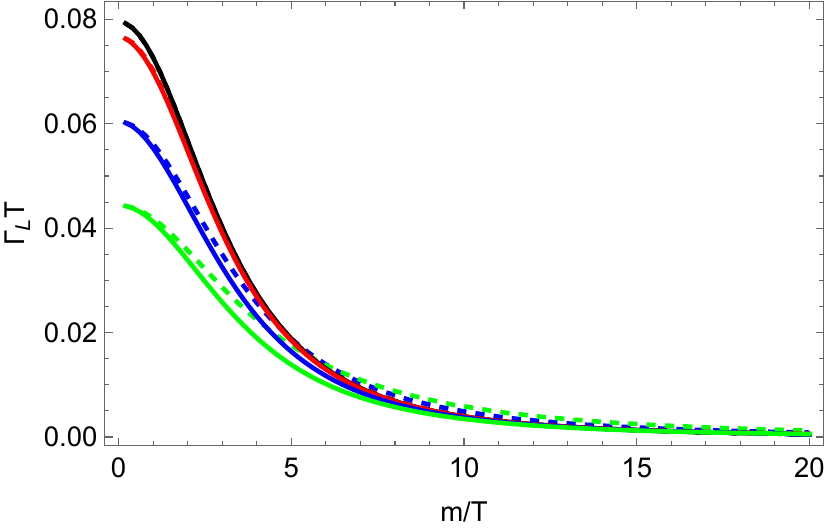}
    \caption{The coefficients in the longitudinal sound modes as the function of $m/T$ for $\mu/T =\{0,1,3,5\}$ (black, red, blue, green). \textbf{Left:} Longitudinal sound velocity $v_L$. \textbf{Right:} Dimensionless longitudinal sound attenuation $\Gamma_LT$.}
    \label{longisound}
\end{figure}

\begin{figure}[htbp]
    \centering
\includegraphics[width=0.6
    \linewidth]{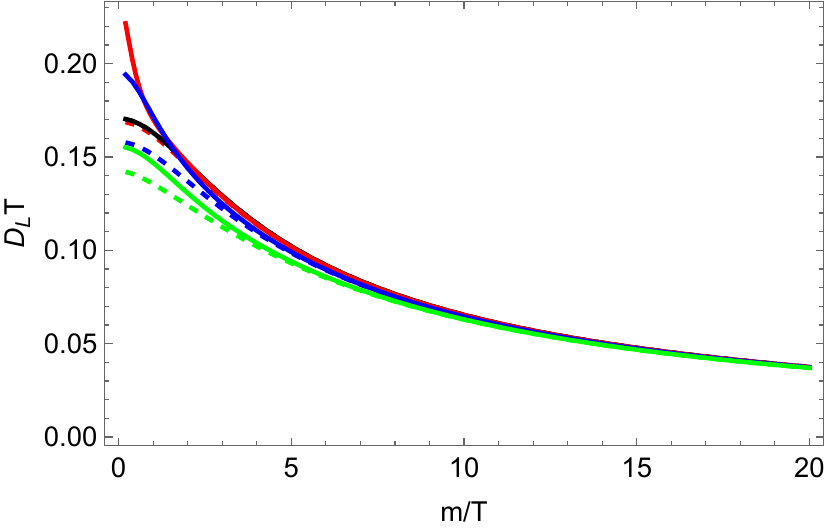}
    \caption{The dimensionless longitudinal diffusivity $D_L T$ as the function of $m/T$ for $\mu/T =\{0,1,3,5\}$ (black, red, blue, green). }
    \label{longicrystal}
\end{figure}

To analyze how the subdiffusive mode influences the dynamics of the phonons and the longitudinal diffusion in the charged case, we compare the full results with those in the absence of $\delta A_a^I$, which is crucial to generate the subdiffusive mode. As is illustrated in FIG.~\ref{transsound} and FIG.~\ref{longisound}, this mode affects the dynamics of the transverse and the longitudinal phonons only via their dissipative parts, leading to a suppression of the sound attenuations. From the hydrodynamic prediction, phonon velocities can be expressed as
\begin{equation}
    v_T=\sqrt{\frac{G}{\chi_{\pi\pi}}},\quad v_L=\sqrt{\frac{G+K}{\chi_{\pi\pi}}},\quad \text{with}\,\,\chi_{\pi\pi}=\epsilon+\mathcal{P}=\frac{3}{2}\epsilon,
\end{equation}
 where $\chi_{\pi\pi}$, $G$, $K$, and $\mathcal{P}$ are the momentum susceptibility, shear modulus, bulk modulus, and pressure, respectively. In addition, $\epsilon$ denotes the energy density which can be calculated by
\begin{equation}
    \epsilon=\frac{1}{u_h^3}\bigg(1+m^2\frac{u_h^{2N}}{2N-3}+\frac{\mu^2u^2_h}{2}\bigg).
\end{equation}
The modulus $G$ and $K$ can be extracted from the real parts of the stress tensor's retarded Green's function $\mathcal{G}^{R}_{T_{ij},T_{kl}}$, via the Kubo formulas. Since $\delta A_a^I$ does not couple to the metric $\delta g_{ij}$ in the EOMs, $\mathcal{G}^{R}_{T_{ij},T_{kl}}$ remains unchanged when $\delta A_a^I$ is turned off. Consequently, the elastic moduli and the two phonon velocities are NOT affected.\footnote{Since the boundary solid is conformal, it possesses a constraint relating the sound speeds, $v_L^2=v_T^2+\frac{1}{2}$ \cite{Esposito:2017qpj}. Given our finding that the transverse phonon velocity is unaltered in the presence of the subdiffusive mode, it is straightforward to show that the longitudinal velocity is also unaltered. However, it is not clear whether this is correct for non-conformal systems.} For the longitudinal diffusive mode, as shown in FIG.~\ref{longicrystal}, the existence of $\delta A_a^I$ enhances the diffusivity, in contrast to its effect on the sound attenuations.

\begin{figure}[htbp]
    \centering
\includegraphics[width=0.6
    \linewidth]{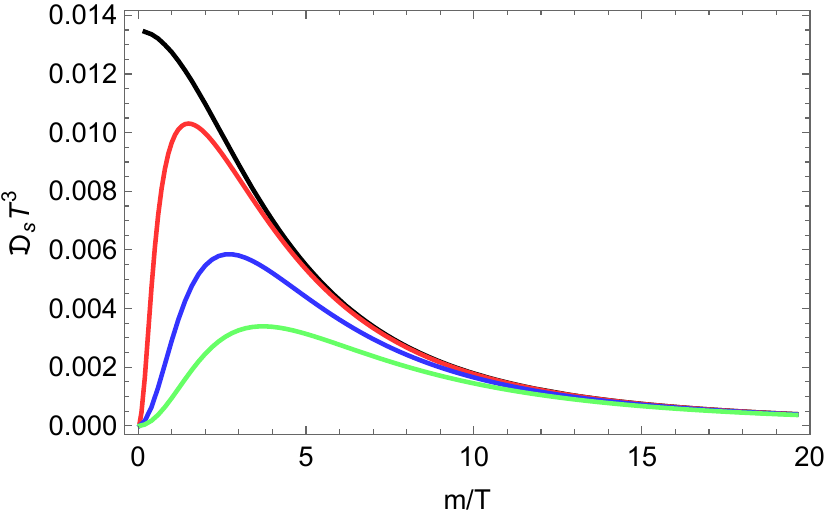}
    \caption{The dimensionless subdiffusivity $\mathfrak{D}_s T^3$ as the function of $m/T$ for $\mu/T =\{0,1,3,5\}$ (black, red, blue, green).}
    \label{longisub}
\end{figure}

In addition, the subdiffusivity is significantly altered by the charged solid background. FIG.~\ref{longisub} displays the dimensionless subdiffusivity $\mathfrak{D}_s T^3$, as a function of $m/T$. It shows that $\mathfrak{D}_s T^3$ exhibits a non-monotonic behavior for the finite density case, peaking at an intermediate temperature. Increasing $\mu/T$ causes this peak to shift towards lower temperatures and decrease in magnitude. When $\mu=0$, the peak is located at zero $m/T$. As a result, $\mathfrak{D}_s T^3$ only decreases monotonically with increasing $m/T$. 

\subsection{Pseudo-spontaneous breaking of translations}\label{section32}

In this subsection, we consider pseudo-spontaneous breaking for translational symmetry. The form of the potential $V(X)$ is set as \cite{Baggioli:2014roa}
\begin{equation}
    V(X)=X +\beta X^N,\quad \text{with}\,\,N>\frac{5}{2}.
\end{equation}
In this case, the boundary momentum is relaxed due to the linear axion term. The new parameter $\beta$ controls the amount of the spontaneous versus explicit breaking. For the pseudo-spontaneous breaking of translations, we require that $\beta \gg 1 $ so that the phonons are just slightly gapped. As in the previous subsection, we fix $N=5$ for the plots shown below.
\begin{figure}[htbp]
    \centering
\includegraphics[width=0.48
    \linewidth]{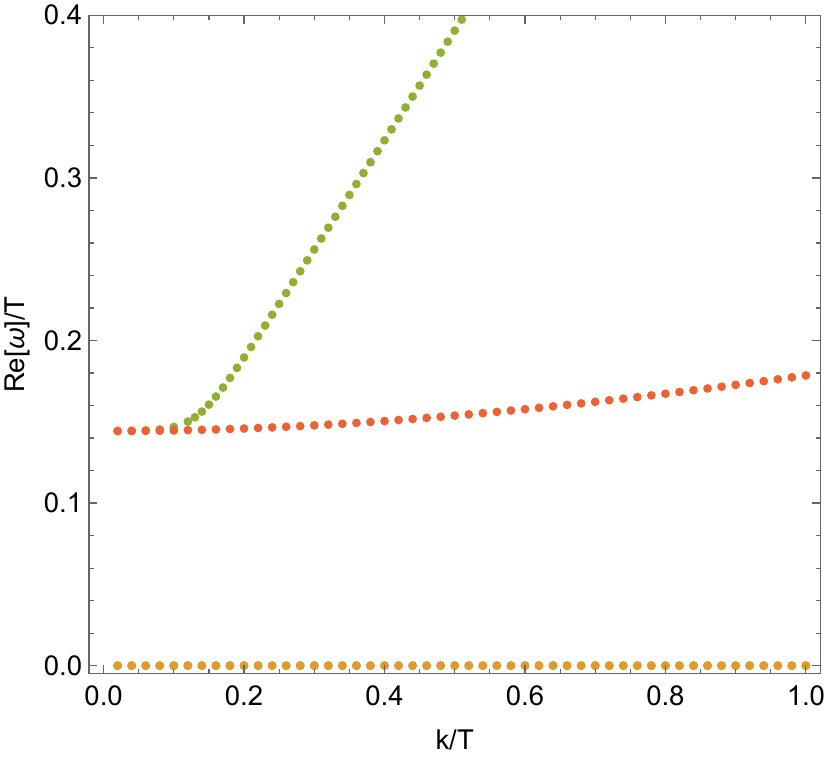}
\includegraphics[width=0.48
    \linewidth]{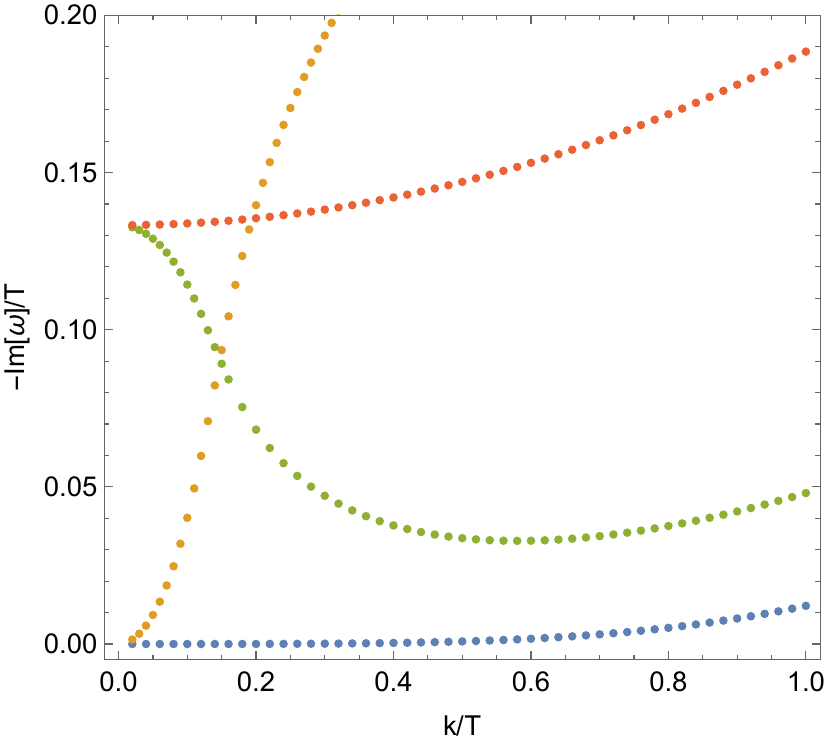}
    \caption{Dispersion relations of (quasi-)hydrodynamic modes in the pseudo-spontaneous breaking case. \textbf{Left:} Real part. \textbf{Right:} Imaginary part. One can clearly see that the longitudinal and transverse phonons (green and red dots) are gapped due to the explicit breaking of translations, while there still exist a diffusive mode (orange dots) and the gapless subdiffusive mode (blue dots) in the longitudinal channel. Here, we fix $\mu/T=0.05,\ m/T=0.06,\ \beta=50$.}
    \label{pseudofull}
\end{figure}
\begin{figure}[htbp]
    \centering
\includegraphics[width=0.7
    \linewidth]{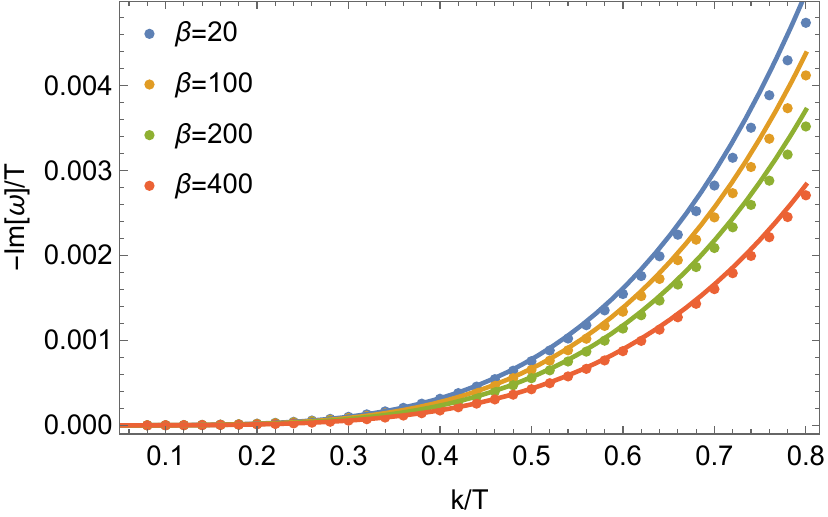}
    \caption{Dispersion relations of the subdiffusive mode for different values of $\beta$, where the dots are the numerical data points and the solid lines represent the hydrodynamic fitting $\omega \sim -ik^4$. Here, we fix $\mu/T=1,\ m/T=0.2$.}
    \label{pseudosub}
\end{figure}
 
 The dispersion relations of the collective modes are displayed in FIG.~\ref{pseudofull}. As we can see, while the phonons acquire a small mass gap due to the slowly relaxed momentum, the subdiffusive mode still remains gapless. The gapless nature of this mode in the pseudo-spontaneous breaking pattern can be viewed as a direct consequence of its protection by the crystal-dipole conservation. Finally, FIG.~\ref{pseudosub} shows how the subdiffusivity decreases as $\beta$ increases.

\section{Discussion and outlook}\label{section4}
In this work, we investigate the collective dynamics in a holographic model of fractonic solids by computing the low-lying QNMs in the bulk. Our results show that, for a fractonic solid system, the combination of charge conservation and crystal-dipole conservation generates a subdiffusive collective mode, which supplants the usual picture of charge diffusion in systems with a single monopole charge conservation. We explicitly verify the existence of this mode, as anticipated in \cite{Jain:2024ngx}, and reveal that it coexists with the acoustic phonons and the crystal diffusion. Furthermore, it influences the dynamics of the other modes only through their dissipative sectors. The coupling between this hydrodynamic mode and the solid background results in a non-monotonic temperature dependence of the subdiffusivity. Another result of our work concerns the stability of this excitation. It is found that this mode remains gapless even when translations are explicitly broken, reflecting the robustness of such excitations in dirty solid systems.

Despite this work marking a step forward in our understanding of fractonic solids, a comprehensive picture of fractonic phases remains far from complete. For instance, an exploration of the non-hydrodynamic modes would be necessary to understand how the system behaves in the high-energy regime. This model can also be generalized to include the conservation of multipole moments up to the order $n$, which is expected to yield a low-energy dispersion relation of the form $\omega \sim -i\, k^{2(n+1)}$. Another appealing direction is to extend this holographic approach to other exotic phases, such as fractonic superfluids and supersolids \cite{Armas:2023ouk,Yuan:2019geh,Jain:2024ngx,Jain:2023nbf,Jain:2024kri,Glodkowski:2025tnv,Stahl:2023prt,Glodkowski:2024ova}. Although the hydrodynamics for such phases have been constructed, the corresponding holographic descriptions are still largely unexplored. Constructing the holographic models would deepen our understanding of the relationship between multipole symmetries and superfluidity, particularly beyond the hydrodynamic regime. Finally, hydrodynamic works \cite{Grosvenor:2021rrt,Glorioso:2021bif,Glorioso:2023chm,Osborne:2021mej,Glodkowski:2022xje} have shown that fluids conserving both spatial-dipole and momentum exhibit a subdiffusive mode and magnon-like Goldstones with $\omega\sim k^2$ dispersion, which motivates the question of how these modes behave when momentum is not conserved. We leave these for future work.

\subsection*{Acknowledgments} 
We would like to thank Matteo Baggioli for fruitful discussions on fractonic phases. This work is supported by the National Natural Science Foundation of China (NSFC) under Grants No.12275038 and No.12475047.

\bibliographystyle{apsrev4-2}
\bibliography{Fractonicsolids}

\begin{thebibliography}{58}%
\makeatletter
\providecommand \@ifxundefined [1]{%
 \@ifx{#1\undefined}
}%
\providecommand \@ifnum [1]{%
 \ifnum #1\expandafter \@firstoftwo
 \else \expandafter \@secondoftwo
 \fi
}%
\providecommand \@ifx [1]{%
 \ifx #1\expandafter \@firstoftwo
 \else \expandafter \@secondoftwo
 \fi
}%
\providecommand \natexlab [1]{#1}%
\providecommand \enquote  [1]{``#1''}%
\providecommand \bibnamefont  [1]{#1}%
\providecommand \bibfnamefont [1]{#1}%
\providecommand \citenamefont [1]{#1}%
\providecommand \href@noop [0]{\@secondoftwo}%
\providecommand \href [0]{\begingroup \@sanitize@url \@href}%
\providecommand \@href[1]{\@@startlink{#1}\@@href}%
\providecommand \@@href[1]{\endgroup#1\@@endlink}%
\providecommand \@sanitize@url [0]{\catcode `\\12\catcode `\$12\catcode `\&12\catcode `\#12\catcode `\^12\catcode `\_12\catcode `\%12\relax}%
\providecommand \@@startlink[1]{}%
\providecommand \@@endlink[0]{}%
\providecommand \url  [0]{\begingroup\@sanitize@url \@url }%
\providecommand \@url [1]{\endgroup\@href {#1}{\urlprefix }}%
\providecommand \urlprefix  [0]{URL }%
\providecommand \Eprint [0]{\href }%
\providecommand \doibase [0]{https://doi.org/}%
\providecommand \selectlanguage [0]{\@gobble}%
\providecommand \bibinfo  [0]{\@secondoftwo}%
\providecommand \bibfield  [0]{\@secondoftwo}%
\providecommand \translation [1]{[#1]}%
\providecommand \BibitemOpen [0]{}%
\providecommand \bibitemStop [0]{}%
\providecommand \bibitemNoStop [0]{.\EOS\space}%
\providecommand \EOS [0]{\spacefactor3000\relax}%
\providecommand \BibitemShut  [1]{\csname bibitem#1\endcsname}%
\let\auto@bib@innerbib\@empty
\bibitem [{\citenamefont {Chamon}(2005)}]{Chamon:2004lew}%
  \BibitemOpen
  \bibfield  {author} {\bibinfo {author} {\bibfnamefont {C.}~\bibnamefont {Chamon}},\ }\href {https://doi.org/10.1103/physrevlett.94.040402} {\bibfield  {journal} {\bibinfo  {journal} {Phys. Rev. Lett.}\ }\textbf {\bibinfo {volume} {94}},\ \bibinfo {pages} {040402} (\bibinfo {year} {2005})},\ \Eprint {https://arxiv.org/abs/cond-mat/0404182} {arXiv:cond-mat/0404182} \BibitemShut {NoStop}%
\bibitem [{\citenamefont {Dubovsky}\ \emph {et~al.}(2006)\citenamefont {Dubovsky}, \citenamefont {Gregoire}, \citenamefont {Nicolis},\ and\ \citenamefont {Rattazzi}}]{Dubovsky:2005xd}%
  \BibitemOpen
  \bibfield  {author} {\bibinfo {author} {\bibfnamefont {S.}~\bibnamefont {Dubovsky}}, \bibinfo {author} {\bibfnamefont {T.}~\bibnamefont {Gregoire}}, \bibinfo {author} {\bibfnamefont {A.}~\bibnamefont {Nicolis}},\ and\ \bibinfo {author} {\bibfnamefont {R.}~\bibnamefont {Rattazzi}},\ }\href {https://doi.org/10.1088/1126-6708/2006/03/025} {\bibfield  {journal} {\bibinfo  {journal} {JHEP}\ }\textbf {\bibinfo {volume} {03}},\ \bibinfo {pages} {025}},\ \Eprint {https://arxiv.org/abs/hep-th/0512260} {arXiv:hep-th/0512260} \BibitemShut {NoStop}%
\bibitem [{\citenamefont {Haah}(2011)}]{Haah:2011drr}%
  \BibitemOpen
  \bibfield  {author} {\bibinfo {author} {\bibfnamefont {J.}~\bibnamefont {Haah}},\ }\href {https://doi.org/10.1103/physreva.83.042330} {\bibfield  {journal} {\bibinfo  {journal} {Phys. Rev. A}\ }\textbf {\bibinfo {volume} {83}},\ \bibinfo {pages} {042330} (\bibinfo {year} {2011})},\ \Eprint {https://arxiv.org/abs/1101.1962} {arXiv:1101.1962 [quant-ph]} \BibitemShut {NoStop}%
\bibitem [{\citenamefont {Vijay}\ \emph {et~al.}(2015)\citenamefont {Vijay}, \citenamefont {Haah},\ and\ \citenamefont {Fu}}]{Vijay:2015mka}%
  \BibitemOpen
  \bibfield  {author} {\bibinfo {author} {\bibfnamefont {S.}~\bibnamefont {Vijay}}, \bibinfo {author} {\bibfnamefont {J.}~\bibnamefont {Haah}},\ and\ \bibinfo {author} {\bibfnamefont {L.}~\bibnamefont {Fu}},\ }\href {https://doi.org/10.1103/PhysRevB.92.235136} {\bibfield  {journal} {\bibinfo  {journal} {Phys. Rev. B}\ }\textbf {\bibinfo {volume} {92}},\ \bibinfo {pages} {235136} (\bibinfo {year} {2015})},\ \Eprint {https://arxiv.org/abs/1505.02576} {arXiv:1505.02576 [cond-mat.str-el]} \BibitemShut {NoStop}%
\bibitem [{\citenamefont {Vijay}\ \emph {et~al.}(2016)\citenamefont {Vijay}, \citenamefont {Haah},\ and\ \citenamefont {Fu}}]{Vijay:2016phm}%
  \BibitemOpen
  \bibfield  {author} {\bibinfo {author} {\bibfnamefont {S.}~\bibnamefont {Vijay}}, \bibinfo {author} {\bibfnamefont {J.}~\bibnamefont {Haah}},\ and\ \bibinfo {author} {\bibfnamefont {L.}~\bibnamefont {Fu}},\ }\href {https://doi.org/10.1103/PhysRevB.94.235157} {\bibfield  {journal} {\bibinfo  {journal} {Phys. Rev. B}\ }\textbf {\bibinfo {volume} {94}},\ \bibinfo {pages} {235157} (\bibinfo {year} {2016})},\ \Eprint {https://arxiv.org/abs/1603.04442} {arXiv:1603.04442 [cond-mat.str-el]} \BibitemShut {NoStop}%
\bibitem [{\citenamefont {Qi}\ \emph {et~al.}(2023)\citenamefont {Qi}, \citenamefont {Hart}, \citenamefont {Friedman}, \citenamefont {Nandkishore},\ and\ \citenamefont {Lucas}}]{Qi:2022seq}%
  \BibitemOpen
  \bibfield  {author} {\bibinfo {author} {\bibfnamefont {M.}~\bibnamefont {Qi}}, \bibinfo {author} {\bibfnamefont {O.}~\bibnamefont {Hart}}, \bibinfo {author} {\bibfnamefont {A.~J.}\ \bibnamefont {Friedman}}, \bibinfo {author} {\bibfnamefont {R.}~\bibnamefont {Nandkishore}},\ and\ \bibinfo {author} {\bibfnamefont {A.}~\bibnamefont {Lucas}},\ }\href {https://doi.org/10.21468/SciPostPhys.14.3.029} {\bibfield  {journal} {\bibinfo  {journal} {SciPost Phys.}\ }\textbf {\bibinfo {volume} {14}},\ \bibinfo {pages} {029} (\bibinfo {year} {2023})},\ \Eprint {https://arxiv.org/abs/2205.05695} {arXiv:2205.05695 [cond-mat.str-el]} \BibitemShut {NoStop}%
\bibitem [{\citenamefont {Jain}\ and\ \citenamefont {Jensen}(2022)}]{Jain:2021ibh}%
  \BibitemOpen
  \bibfield  {author} {\bibinfo {author} {\bibfnamefont {A.}~\bibnamefont {Jain}}\ and\ \bibinfo {author} {\bibfnamefont {K.}~\bibnamefont {Jensen}},\ }\href {https://doi.org/10.21468/SciPostPhys.12.4.142} {\bibfield  {journal} {\bibinfo  {journal} {SciPost Phys.}\ }\textbf {\bibinfo {volume} {12}},\ \bibinfo {pages} {142} (\bibinfo {year} {2022})},\ \Eprint {https://arxiv.org/abs/2111.03973} {arXiv:2111.03973 [hep-th]} \BibitemShut {NoStop}%
\bibitem [{\citenamefont {Bidussi}\ \emph {et~al.}(2022)\citenamefont {Bidussi}, \citenamefont {Hartong}, \citenamefont {Have}, \citenamefont {Musaeus},\ and\ \citenamefont {Prohazka}}]{Bidussi:2021nmp}%
  \BibitemOpen
  \bibfield  {author} {\bibinfo {author} {\bibfnamefont {L.}~\bibnamefont {Bidussi}}, \bibinfo {author} {\bibfnamefont {J.}~\bibnamefont {Hartong}}, \bibinfo {author} {\bibfnamefont {E.}~\bibnamefont {Have}}, \bibinfo {author} {\bibfnamefont {J.}~\bibnamefont {Musaeus}},\ and\ \bibinfo {author} {\bibfnamefont {S.}~\bibnamefont {Prohazka}},\ }\href {https://doi.org/10.21468/SciPostPhys.12.6.205} {\bibfield  {journal} {\bibinfo  {journal} {SciPost Phys.}\ }\textbf {\bibinfo {volume} {12}},\ \bibinfo {pages} {205} (\bibinfo {year} {2022})},\ \Eprint {https://arxiv.org/abs/2111.03668} {arXiv:2111.03668 [hep-th]} \BibitemShut {NoStop}%
\bibitem [{\citenamefont {Khlopov}(1981)}]{Khlopov:1981wm}%
  \BibitemOpen
  \bibfield  {author} {\bibinfo {author} {\bibfnamefont {M.~Y.}\ \bibnamefont {Khlopov}},\ }\href@noop {} {\bibfield  {journal} {\bibinfo  {journal} {Pisma Zh. Eksp. Teor. Fiz.}\ }\textbf {\bibinfo {volume} {33}},\ \bibinfo {pages} {170} (\bibinfo {year} {1981})}\BibitemShut {NoStop}%
\bibitem [{\citenamefont {Seiberg}\ and\ \citenamefont {Shao}(2020)}]{Seiberg:2020wsg}%
  \BibitemOpen
  \bibfield  {author} {\bibinfo {author} {\bibfnamefont {N.}~\bibnamefont {Seiberg}}\ and\ \bibinfo {author} {\bibfnamefont {S.-H.}\ \bibnamefont {Shao}},\ }\href {https://doi.org/10.21468/SciPostPhys.9.4.046} {\bibfield  {journal} {\bibinfo  {journal} {SciPost Phys.}\ }\textbf {\bibinfo {volume} {9}},\ \bibinfo {pages} {046} (\bibinfo {year} {2020})},\ \Eprint {https://arxiv.org/abs/2004.00015} {arXiv:2004.00015 [cond-mat.str-el]} \BibitemShut {NoStop}%
\bibitem [{\citenamefont {You}\ \emph {et~al.}(2021)\citenamefont {You}, \citenamefont {Bibo}, \citenamefont {Hughes},\ and\ \citenamefont {Pollmann}}]{You:2021tmm}%
  \BibitemOpen
  \bibfield  {author} {\bibinfo {author} {\bibfnamefont {Y.}~\bibnamefont {You}}, \bibinfo {author} {\bibfnamefont {J.}~\bibnamefont {Bibo}}, \bibinfo {author} {\bibfnamefont {T.~L.}\ \bibnamefont {Hughes}},\ and\ \bibinfo {author} {\bibfnamefont {F.}~\bibnamefont {Pollmann}},\ }\href@noop {} {\bibinfo {title} {{Fractonic critical point proximate to a higher-order topological insulator: How does UV blend with IR?}}} (\bibinfo {year} {2021}),\ \Eprint {https://arxiv.org/abs/2101.01724} {arXiv:2101.01724 [cond-mat.str-el]} \BibitemShut {NoStop}%
\bibitem [{\citenamefont {Gorantla}\ \emph {et~al.}(2021)\citenamefont {Gorantla}, \citenamefont {Lam}, \citenamefont {Seiberg},\ and\ \citenamefont {Shao}}]{Gorantla:2021bda}%
  \BibitemOpen
  \bibfield  {author} {\bibinfo {author} {\bibfnamefont {P.}~\bibnamefont {Gorantla}}, \bibinfo {author} {\bibfnamefont {H.~T.}\ \bibnamefont {Lam}}, \bibinfo {author} {\bibfnamefont {N.}~\bibnamefont {Seiberg}},\ and\ \bibinfo {author} {\bibfnamefont {S.-H.}\ \bibnamefont {Shao}},\ }\href {https://doi.org/10.1103/PhysRevB.104.235116} {\bibfield  {journal} {\bibinfo  {journal} {Phys. Rev. B}\ }\textbf {\bibinfo {volume} {104}},\ \bibinfo {pages} {235116} (\bibinfo {year} {2021})},\ \Eprint {https://arxiv.org/abs/2108.00020} {arXiv:2108.00020 [cond-mat.str-el]} \BibitemShut {NoStop}%
\bibitem [{\citenamefont {Pretko}(2018)}]{Pretko:2018jbi}%
  \BibitemOpen
  \bibfield  {author} {\bibinfo {author} {\bibfnamefont {M.}~\bibnamefont {Pretko}},\ }\href {https://doi.org/10.1103/PhysRevB.98.115134} {\bibfield  {journal} {\bibinfo  {journal} {Phys. Rev. B}\ }\textbf {\bibinfo {volume} {98}},\ \bibinfo {pages} {115134} (\bibinfo {year} {2018})},\ \Eprint {https://arxiv.org/abs/1807.11479} {arXiv:1807.11479 [cond-mat.str-el]} \BibitemShut {NoStop}%
\bibitem [{\citenamefont {Pretko}\ \emph {et~al.}(2020)\citenamefont {Pretko}, \citenamefont {Chen},\ and\ \citenamefont {You}}]{Pretko:2020cko}%
  \BibitemOpen
  \bibfield  {author} {\bibinfo {author} {\bibfnamefont {M.}~\bibnamefont {Pretko}}, \bibinfo {author} {\bibfnamefont {X.}~\bibnamefont {Chen}},\ and\ \bibinfo {author} {\bibfnamefont {Y.}~\bibnamefont {You}},\ }\href {https://doi.org/10.1142/S0217751X20300033} {\bibfield  {journal} {\bibinfo  {journal} {Int. J. Mod. Phys. A}\ }\textbf {\bibinfo {volume} {35}},\ \bibinfo {pages} {2030003} (\bibinfo {year} {2020})},\ \Eprint {https://arxiv.org/abs/2001.01722} {arXiv:2001.01722 [cond-mat.str-el]} \BibitemShut {NoStop}%
\bibitem [{\citenamefont {Prem}\ \emph {et~al.}(2017)\citenamefont {Prem}, \citenamefont {Haah},\ and\ \citenamefont {Nandkishore}}]{Prem:2017qcp}%
  \BibitemOpen
  \bibfield  {author} {\bibinfo {author} {\bibfnamefont {A.}~\bibnamefont {Prem}}, \bibinfo {author} {\bibfnamefont {J.}~\bibnamefont {Haah}},\ and\ \bibinfo {author} {\bibfnamefont {R.}~\bibnamefont {Nandkishore}},\ }\href {https://doi.org/10.1103/PhysRevB.95.155133} {\bibfield  {journal} {\bibinfo  {journal} {Phys. Rev. B}\ }\textbf {\bibinfo {volume} {95}},\ \bibinfo {pages} {155133} (\bibinfo {year} {2017})},\ \Eprint {https://arxiv.org/abs/1702.02952} {arXiv:1702.02952 [cond-mat.stat-mech]} \BibitemShut {NoStop}%
\bibitem [{\citenamefont {Pai}\ \emph {et~al.}(2019)\citenamefont {Pai}, \citenamefont {Pretko},\ and\ \citenamefont {Nandkishore}}]{Pai:2018gyd}%
  \BibitemOpen
  \bibfield  {author} {\bibinfo {author} {\bibfnamefont {S.}~\bibnamefont {Pai}}, \bibinfo {author} {\bibfnamefont {M.}~\bibnamefont {Pretko}},\ and\ \bibinfo {author} {\bibfnamefont {R.~M.}\ \bibnamefont {Nandkishore}},\ }\href {https://doi.org/10.1103/PhysRevX.9.021003} {\bibfield  {journal} {\bibinfo  {journal} {Phys. Rev. X}\ }\textbf {\bibinfo {volume} {9}},\ \bibinfo {pages} {021003} (\bibinfo {year} {2019})},\ \bibinfo {note} {[Erratum: Phys.Rev.X 9, 049901 (2019)]},\ \Eprint {https://arxiv.org/abs/1807.09776} {arXiv:1807.09776 [cond-mat.stat-mech]} \BibitemShut {NoStop}%
\bibitem [{\citenamefont {Brown}\ and\ \citenamefont {Williamson}(2020)}]{Brown:2019hxw}%
  \BibitemOpen
  \bibfield  {author} {\bibinfo {author} {\bibfnamefont {B.~J.}\ \bibnamefont {Brown}}\ and\ \bibinfo {author} {\bibfnamefont {D.~J.}\ \bibnamefont {Williamson}},\ }\href {https://doi.org/10.1103/PhysRevResearch.2.013303} {\bibfield  {journal} {\bibinfo  {journal} {Phys. Rev. Res.}\ }\textbf {\bibinfo {volume} {2}},\ \bibinfo {pages} {013303} (\bibinfo {year} {2020})},\ \Eprint {https://arxiv.org/abs/1901.08061} {arXiv:1901.08061 [quant-ph]} \BibitemShut {NoStop}%
\bibitem [{\citenamefont {Terhal}(2015)}]{Terhal:2013vbm}%
  \BibitemOpen
  \bibfield  {author} {\bibinfo {author} {\bibfnamefont {B.~M.}\ \bibnamefont {Terhal}},\ }\href {https://doi.org/10.1103/RevModPhys.87.307} {\bibfield  {journal} {\bibinfo  {journal} {Rev. Mod. Phys.}\ }\textbf {\bibinfo {volume} {87}},\ \bibinfo {pages} {307} (\bibinfo {year} {2015})},\ \Eprint {https://arxiv.org/abs/1302.3428} {arXiv:1302.3428 [quant-ph]} \BibitemShut {NoStop}%
\bibitem [{\citenamefont {Bravyi}\ and\ \citenamefont {Haah}(2013)}]{Bravyi:2011faf}%
  \BibitemOpen
  \bibfield  {author} {\bibinfo {author} {\bibfnamefont {S.}~\bibnamefont {Bravyi}}\ and\ \bibinfo {author} {\bibfnamefont {J.}~\bibnamefont {Haah}},\ }\href {https://doi.org/10.1103/PhysRevLett.111.200501} {\bibfield  {journal} {\bibinfo  {journal} {Phys. Rev. Lett.}\ }\textbf {\bibinfo {volume} {111}},\ \bibinfo {pages} {200501} (\bibinfo {year} {2013})},\ \Eprint {https://arxiv.org/abs/1112.3252} {arXiv:1112.3252 [quant-ph]} \BibitemShut {NoStop}%
\bibitem [{\citenamefont {Gromov}\ and\ \citenamefont {Radzihovsky}(2024)}]{Gromov:2022cxa}%
  \BibitemOpen
  \bibfield  {author} {\bibinfo {author} {\bibfnamefont {A.}~\bibnamefont {Gromov}}\ and\ \bibinfo {author} {\bibfnamefont {L.}~\bibnamefont {Radzihovsky}},\ }\href {https://doi.org/10.1103/RevModPhys.96.011001} {\bibfield  {journal} {\bibinfo  {journal} {Rev. Mod. Phys.}\ }\textbf {\bibinfo {volume} {96}},\ \bibinfo {pages} {011001} (\bibinfo {year} {2024})},\ \Eprint {https://arxiv.org/abs/2211.05130} {arXiv:2211.05130 [cond-mat.str-el]} \BibitemShut {NoStop}%
\bibitem [{\citenamefont {Nandkishore}\ and\ \citenamefont {Hermele}(2019)}]{Nandkishore:2018sel}%
  \BibitemOpen
  \bibfield  {author} {\bibinfo {author} {\bibfnamefont {R.~M.}\ \bibnamefont {Nandkishore}}\ and\ \bibinfo {author} {\bibfnamefont {M.}~\bibnamefont {Hermele}},\ }\href {https://doi.org/10.1146/annurev-conmatphys-031218-013604} {\bibfield  {journal} {\bibinfo  {journal} {Ann. Rev. Condensed Matter Phys.}\ }\textbf {\bibinfo {volume} {10}},\ \bibinfo {pages} {295} (\bibinfo {year} {2019})},\ \Eprint {https://arxiv.org/abs/1803.11196} {arXiv:1803.11196 [cond-mat.str-el]} \BibitemShut {NoStop}%
\bibitem [{\citenamefont {Gromov}\ \emph {et~al.}(2020)\citenamefont {Gromov}, \citenamefont {Lucas},\ and\ \citenamefont {Nandkishore}}]{Gromov:2020yoc}%
  \BibitemOpen
  \bibfield  {author} {\bibinfo {author} {\bibfnamefont {A.}~\bibnamefont {Gromov}}, \bibinfo {author} {\bibfnamefont {A.}~\bibnamefont {Lucas}},\ and\ \bibinfo {author} {\bibfnamefont {R.~M.}\ \bibnamefont {Nandkishore}},\ }\href {https://doi.org/10.1103/PhysRevResearch.2.033124} {\bibfield  {journal} {\bibinfo  {journal} {Phys. Rev. Res.}\ }\textbf {\bibinfo {volume} {2}},\ \bibinfo {pages} {033124} (\bibinfo {year} {2020})},\ \Eprint {https://arxiv.org/abs/2003.09429} {arXiv:2003.09429 [cond-mat.str-el]} \BibitemShut {NoStop}%
\bibitem [{\citenamefont {Grosvenor}\ \emph {et~al.}(2021)\citenamefont {Grosvenor}, \citenamefont {Hoyos}, \citenamefont {Pe{\~n}a-Ben{\'\i}tez},\ and\ \citenamefont {Sur{\'o}wka}}]{Grosvenor:2021rrt}%
  \BibitemOpen
  \bibfield  {author} {\bibinfo {author} {\bibfnamefont {K.~T.}\ \bibnamefont {Grosvenor}}, \bibinfo {author} {\bibfnamefont {C.}~\bibnamefont {Hoyos}}, \bibinfo {author} {\bibfnamefont {F.}~\bibnamefont {Pe{\~n}a-Ben{\'\i}tez}},\ and\ \bibinfo {author} {\bibfnamefont {P.}~\bibnamefont {Sur{\'o}wka}},\ }\href {https://doi.org/10.1103/PhysRevResearch.3.043186} {\bibfield  {journal} {\bibinfo  {journal} {Phys. Rev. Res.}\ }\textbf {\bibinfo {volume} {3}},\ \bibinfo {pages} {043186} (\bibinfo {year} {2021})},\ \Eprint {https://arxiv.org/abs/2105.01084} {arXiv:2105.01084 [cond-mat.str-el]} \BibitemShut {NoStop}%
\bibitem [{\citenamefont {Glorioso}\ \emph {et~al.}(2022)\citenamefont {Glorioso}, \citenamefont {Guo}, \citenamefont {Rodriguez-Nieva},\ and\ \citenamefont {Lucas}}]{Glorioso:2021bif}%
  \BibitemOpen
  \bibfield  {author} {\bibinfo {author} {\bibfnamefont {P.}~\bibnamefont {Glorioso}}, \bibinfo {author} {\bibfnamefont {J.}~\bibnamefont {Guo}}, \bibinfo {author} {\bibfnamefont {J.~F.}\ \bibnamefont {Rodriguez-Nieva}},\ and\ \bibinfo {author} {\bibfnamefont {A.}~\bibnamefont {Lucas}},\ }\href {https://doi.org/10.1038/s41567-022-01631-x} {\bibfield  {journal} {\bibinfo  {journal} {Nature Phys.}\ }\textbf {\bibinfo {volume} {18}},\ \bibinfo {pages} {912} (\bibinfo {year} {2022})},\ \Eprint {https://arxiv.org/abs/2105.13365} {arXiv:2105.13365 [cond-mat.str-el]} \BibitemShut {NoStop}%
\bibitem [{\citenamefont {Burchards}\ \emph {et~al.}(2022)\citenamefont {Burchards}, \citenamefont {Feldmeier}, \citenamefont {Schuckert},\ and\ \citenamefont {Knap}}]{Burchards:2022lqr}%
  \BibitemOpen
  \bibfield  {author} {\bibinfo {author} {\bibfnamefont {A.~G.}\ \bibnamefont {Burchards}}, \bibinfo {author} {\bibfnamefont {J.}~\bibnamefont {Feldmeier}}, \bibinfo {author} {\bibfnamefont {A.}~\bibnamefont {Schuckert}},\ and\ \bibinfo {author} {\bibfnamefont {M.}~\bibnamefont {Knap}},\ }\href {https://doi.org/10.1103/PhysRevB.105.205127} {\bibfield  {journal} {\bibinfo  {journal} {Phys. Rev. B}\ }\textbf {\bibinfo {volume} {105}},\ \bibinfo {pages} {205127} (\bibinfo {year} {2022})},\ \Eprint {https://arxiv.org/abs/2201.08852} {arXiv:2201.08852 [cond-mat.quant-gas]} \BibitemShut {NoStop}%
\bibitem [{\citenamefont {Guo}\ \emph {et~al.}(2022)\citenamefont {Guo}, \citenamefont {Glorioso},\ and\ \citenamefont {Lucas}}]{Guo:2022ixk}%
  \BibitemOpen
  \bibfield  {author} {\bibinfo {author} {\bibfnamefont {J.}~\bibnamefont {Guo}}, \bibinfo {author} {\bibfnamefont {P.}~\bibnamefont {Glorioso}},\ and\ \bibinfo {author} {\bibfnamefont {A.}~\bibnamefont {Lucas}},\ }\href {https://doi.org/10.1103/PhysRevLett.129.150603} {\bibfield  {journal} {\bibinfo  {journal} {Phys. Rev. Lett.}\ }\textbf {\bibinfo {volume} {129}},\ \bibinfo {pages} {150603} (\bibinfo {year} {2022})},\ \Eprint {https://arxiv.org/abs/2204.06006} {arXiv:2204.06006 [cond-mat.stat-mech]} \BibitemShut {NoStop}%
\bibitem [{\citenamefont {Hart}\ \emph {et~al.}(2022)\citenamefont {Hart}, \citenamefont {Lucas},\ and\ \citenamefont {Nandkishore}}]{Hart_2022}%
  \BibitemOpen
  \bibfield  {author} {\bibinfo {author} {\bibfnamefont {O.}~\bibnamefont {Hart}}, \bibinfo {author} {\bibfnamefont {A.}~\bibnamefont {Lucas}},\ and\ \bibinfo {author} {\bibfnamefont {R.}~\bibnamefont {Nandkishore}},\ }\bibfield  {journal} {\bibinfo  {journal} {Physical Review E}\ }\textbf {\bibinfo {volume} {105}},\ \href {https://doi.org/10.1103/physreve.105.044103} {10.1103/physreve.105.044103} (\bibinfo {year} {2022})\BibitemShut {NoStop}%
\bibitem [{\citenamefont {Glorioso}\ \emph {et~al.}(2023)\citenamefont {Glorioso}, \citenamefont {Huang}, \citenamefont {Guo}, \citenamefont {Rodriguez-Nieva},\ and\ \citenamefont {Lucas}}]{Glorioso:2023chm}%
  \BibitemOpen
  \bibfield  {author} {\bibinfo {author} {\bibfnamefont {P.}~\bibnamefont {Glorioso}}, \bibinfo {author} {\bibfnamefont {X.}~\bibnamefont {Huang}}, \bibinfo {author} {\bibfnamefont {J.}~\bibnamefont {Guo}}, \bibinfo {author} {\bibfnamefont {J.~F.}\ \bibnamefont {Rodriguez-Nieva}},\ and\ \bibinfo {author} {\bibfnamefont {A.}~\bibnamefont {Lucas}},\ }\href {https://doi.org/10.1007/JHEP05(2023)022} {\bibfield  {journal} {\bibinfo  {journal} {JHEP}\ }\textbf {\bibinfo {volume} {05}},\ \bibinfo {pages} {022}},\ \Eprint {https://arxiv.org/abs/2301.02680} {arXiv:2301.02680 [hep-th]} \BibitemShut {NoStop}%
\bibitem [{\citenamefont {Osborne}\ and\ \citenamefont {Lucas}(2022)}]{Osborne:2021mej}%
  \BibitemOpen
  \bibfield  {author} {\bibinfo {author} {\bibfnamefont {A.}~\bibnamefont {Osborne}}\ and\ \bibinfo {author} {\bibfnamefont {A.}~\bibnamefont {Lucas}},\ }\href {https://doi.org/10.1103/PhysRevB.105.024311} {\bibfield  {journal} {\bibinfo  {journal} {Phys. Rev. B}\ }\textbf {\bibinfo {volume} {105}},\ \bibinfo {pages} {024311} (\bibinfo {year} {2022})},\ \Eprint {https://arxiv.org/abs/2111.09323} {arXiv:2111.09323 [cond-mat.stat-mech]} \BibitemShut {NoStop}%
\bibitem [{\citenamefont {G{\l}{\'o}dkowski}\ \emph {et~al.}(2023)\citenamefont {G{\l}{\'o}dkowski}, \citenamefont {Pe{\~n}a-Ben{\'\i}tez},\ and\ \citenamefont {Sur{\'o}wka}}]{Glodkowski:2022xje}%
  \BibitemOpen
  \bibfield  {author} {\bibinfo {author} {\bibfnamefont {A.}~\bibnamefont {G{\l}{\'o}dkowski}}, \bibinfo {author} {\bibfnamefont {F.}~\bibnamefont {Pe{\~n}a-Ben{\'\i}tez}},\ and\ \bibinfo {author} {\bibfnamefont {P.}~\bibnamefont {Sur{\'o}wka}},\ }\href {https://doi.org/10.1103/PhysRevE.107.034142} {\bibfield  {journal} {\bibinfo  {journal} {Phys. Rev. E}\ }\textbf {\bibinfo {volume} {107}},\ \bibinfo {pages} {034142} (\bibinfo {year} {2023})},\ \Eprint {https://arxiv.org/abs/2212.06848} {arXiv:2212.06848 [cond-mat.str-el]} \BibitemShut {NoStop}%
\bibitem [{\citenamefont {Jain}\ \emph {et~al.}(2023)\citenamefont {Jain}, \citenamefont {Jensen}, \citenamefont {Liu},\ and\ \citenamefont {Mefford}}]{Jain:2023nbf}%
  \BibitemOpen
  \bibfield  {author} {\bibinfo {author} {\bibfnamefont {A.}~\bibnamefont {Jain}}, \bibinfo {author} {\bibfnamefont {K.}~\bibnamefont {Jensen}}, \bibinfo {author} {\bibfnamefont {R.}~\bibnamefont {Liu}},\ and\ \bibinfo {author} {\bibfnamefont {E.}~\bibnamefont {Mefford}},\ }\href {https://doi.org/10.1007/JHEP09(2023)184} {\bibfield  {journal} {\bibinfo  {journal} {JHEP}\ }\textbf {\bibinfo {volume} {09}},\ \bibinfo {pages} {184}},\ \Eprint {https://arxiv.org/abs/2304.09852} {arXiv:2304.09852 [hep-th]} \BibitemShut {NoStop}%
\bibitem [{\citenamefont {Jain}\ \emph {et~al.}(2024)\citenamefont {Jain}, \citenamefont {Jensen}, \citenamefont {Liu},\ and\ \citenamefont {Mefford}}]{Jain:2024kri}%
  \BibitemOpen
  \bibfield  {author} {\bibinfo {author} {\bibfnamefont {A.}~\bibnamefont {Jain}}, \bibinfo {author} {\bibfnamefont {K.}~\bibnamefont {Jensen}}, \bibinfo {author} {\bibfnamefont {R.}~\bibnamefont {Liu}},\ and\ \bibinfo {author} {\bibfnamefont {E.}~\bibnamefont {Mefford}},\ }\href {https://doi.org/10.1007/JHEP07(2024)197} {\bibfield  {journal} {\bibinfo  {journal} {JHEP}\ }\textbf {\bibinfo {volume} {07}},\ \bibinfo {pages} {197}},\ \Eprint {https://arxiv.org/abs/2401.16385} {arXiv:2401.16385 [hep-th]} \BibitemShut {NoStop}%
\bibitem [{\citenamefont {Jain}(2024)}]{Jain:2024ngx}%
  \BibitemOpen
  \bibfield  {author} {\bibinfo {author} {\bibfnamefont {A.}~\bibnamefont {Jain}},\ }\href@noop {} {\bibinfo {title} {{Fractonic solids}}} (\bibinfo {year} {2024}),\ \Eprint {https://arxiv.org/abs/2406.07334} {arXiv:2406.07334 [hep-th]} \BibitemShut {NoStop}%
\bibitem [{\citenamefont {Gromov}(2019)}]{Gromov:2018nbv}%
  \BibitemOpen
  \bibfield  {author} {\bibinfo {author} {\bibfnamefont {A.}~\bibnamefont {Gromov}},\ }\href {https://doi.org/10.1103/PhysRevX.9.031035} {\bibfield  {journal} {\bibinfo  {journal} {Phys. Rev. X}\ }\textbf {\bibinfo {volume} {9}},\ \bibinfo {pages} {031035} (\bibinfo {year} {2019})},\ \Eprint {https://arxiv.org/abs/1812.05104} {arXiv:1812.05104 [cond-mat.str-el]} \BibitemShut {NoStop}%
\bibitem [{\citenamefont {Chaikin}\ and\ \citenamefont {Lubensky}(1995)}]{Lubensky}%
  \BibitemOpen
  \bibfield  {author} {\bibinfo {author} {\bibfnamefont {P.~M.}\ \bibnamefont {Chaikin}}\ and\ \bibinfo {author} {\bibfnamefont {T.~C.}\ \bibnamefont {Lubensky}},\ }\href {https://doi.org/10.1017/CBO9780511813467} {\emph {\bibinfo {title} {Principles of Condensed Matter Physics}}}\ (\bibinfo  {publisher} {Cambridge University Press},\ \bibinfo {year} {1995})\BibitemShut {NoStop}%
\bibitem [{\citenamefont {Gr\"uner}(1988)}]{RevModPhys.60.1129}%
  \BibitemOpen
  \bibfield  {author} {\bibinfo {author} {\bibfnamefont {G.}~\bibnamefont {Gr\"uner}},\ }\href {https://doi.org/10.1103/RevModPhys.60.1129} {\bibfield  {journal} {\bibinfo  {journal} {Rev. Mod. Phys.}\ }\textbf {\bibinfo {volume} {60}},\ \bibinfo {pages} {1129} (\bibinfo {year} {1988})}\BibitemShut {NoStop}%
\bibitem [{\citenamefont {Ammon}\ and\ \citenamefont {Erdmenger}(2015)}]{Ammon_Erdmenger_2015}%
  \BibitemOpen
  \bibfield  {author} {\bibinfo {author} {\bibfnamefont {M.}~\bibnamefont {Ammon}}\ and\ \bibinfo {author} {\bibfnamefont {J.}~\bibnamefont {Erdmenger}},\ }\href {https://doi.org/10.1017/CBO9780511846373} {\emph {\bibinfo {title} {Gauge/Gravity Duality: Foundations and Applications}}}\ (\bibinfo  {publisher} {Cambridge University Press},\ \bibinfo {year} {2015})\BibitemShut {NoStop}%
\bibitem [{\citenamefont {Baggioli}(2019)}]{Baggioli:2019rr}%
  \BibitemOpen
  \bibfield  {author} {\bibinfo {author} {\bibfnamefont {M.}~\bibnamefont {Baggioli}},\ }\href {https://doi.org/10.1007/978-3-030-35184-7} {\emph {\bibinfo {title} {{Applied Holography}: {A Practical Mini-Course}}}}\ (\bibinfo  {publisher} {SpringerBriefs in Physics},\ \bibinfo {year} {2019})\BibitemShut {NoStop}%
\bibitem [{\citenamefont {Maldacena}(1998)}]{Maldacena:1997re}%
  \BibitemOpen
  \bibfield  {author} {\bibinfo {author} {\bibfnamefont {J.~M.}\ \bibnamefont {Maldacena}},\ }\href {https://doi.org/10.4310/ATMP.1998.v2.n2.a1} {\bibfield  {journal} {\bibinfo  {journal} {Adv. Theor. Math. Phys.}\ }\textbf {\bibinfo {volume} {2}},\ \bibinfo {pages} {231} (\bibinfo {year} {1998})},\ \Eprint {https://arxiv.org/abs/hep-th/9711200} {arXiv:hep-th/9711200} \BibitemShut {NoStop}%
\bibitem [{\citenamefont {Witten}(1998)}]{Witten:1998qj}%
  \BibitemOpen
  \bibfield  {author} {\bibinfo {author} {\bibfnamefont {E.}~\bibnamefont {Witten}},\ }\href {https://doi.org/10.4310/ATMP.1998.v2.n2.a2} {\bibfield  {journal} {\bibinfo  {journal} {Adv. Theor. Math. Phys.}\ }\textbf {\bibinfo {volume} {2}},\ \bibinfo {pages} {253} (\bibinfo {year} {1998})},\ \Eprint {https://arxiv.org/abs/hep-th/9802150} {arXiv:hep-th/9802150} \BibitemShut {NoStop}%
\bibitem [{\citenamefont {Hartnoll}\ \emph {et~al.}(2018)\citenamefont {Hartnoll}, \citenamefont {Lucas},\ and\ \citenamefont {Sachdev}}]{Hartnoll:2016apf}%
  \BibitemOpen
  \bibfield  {author} {\bibinfo {author} {\bibfnamefont {S.~A.}\ \bibnamefont {Hartnoll}}, \bibinfo {author} {\bibfnamefont {A.}~\bibnamefont {Lucas}},\ and\ \bibinfo {author} {\bibfnamefont {S.}~\bibnamefont {Sachdev}},\ }\href@noop {} {\emph {\bibinfo {title} {Holographic quantum matter}}}\ (\bibinfo  {publisher} {The MIT Press},\ \bibinfo {year} {2018})\BibitemShut {NoStop}%
\bibitem [{\citenamefont {Zaanen}\ \emph {et~al.}(2015)\citenamefont {Zaanen}, \citenamefont {Liu}, \citenamefont {Sun},\ and\ \citenamefont {Schalm}}]{Zaanen_Liu_Sun_Schalm_2015}%
  \BibitemOpen
  \bibfield  {author} {\bibinfo {author} {\bibfnamefont {J.}~\bibnamefont {Zaanen}}, \bibinfo {author} {\bibfnamefont {Y.}~\bibnamefont {Liu}}, \bibinfo {author} {\bibfnamefont {Y.-W.}\ \bibnamefont {Sun}},\ and\ \bibinfo {author} {\bibfnamefont {K.}~\bibnamefont {Schalm}},\ }\href {https://doi.org/10.1017/CBO9781139942492} {\emph {\bibinfo {title} {Holographic Duality in Condensed Matter Physics}}}\ (\bibinfo  {publisher} {Cambridge University Press},\ \bibinfo {year} {2015})\BibitemShut {NoStop}%
\bibitem [{\citenamefont {Baggioli}\ \emph {et~al.}(2021)\citenamefont {Baggioli}, \citenamefont {Kim}, \citenamefont {Li},\ and\ \citenamefont {Li}}]{Baggioli:2021xuv}%
  \BibitemOpen
  \bibfield  {author} {\bibinfo {author} {\bibfnamefont {M.}~\bibnamefont {Baggioli}}, \bibinfo {author} {\bibfnamefont {K.-Y.}\ \bibnamefont {Kim}}, \bibinfo {author} {\bibfnamefont {L.}~\bibnamefont {Li}},\ and\ \bibinfo {author} {\bibfnamefont {W.-J.}\ \bibnamefont {Li}},\ }\href {https://doi.org/10.1007/s11433-021-1681-8} {\bibfield  {journal} {\bibinfo  {journal} {Sci. China Phys. Mech. Astron.}\ }\textbf {\bibinfo {volume} {64}},\ \bibinfo {pages} {270001} (\bibinfo {year} {2021})},\ \Eprint {https://arxiv.org/abs/2101.01892} {arXiv:2101.01892 [hep-th]} \BibitemShut {NoStop}%
\bibitem [{\citenamefont {G{\l}{\'o}dkowski}(2025)}]{Glodkowski:2025tnv}%
  \BibitemOpen
  \bibfield  {author} {\bibinfo {author} {\bibfnamefont {A.}~\bibnamefont {G{\l}{\'o}dkowski}},\ }\href@noop {} {\bibinfo {title} {{A complex scalar field theory for charged fluids, superfluids, and fracton fluids}}} (\bibinfo {year} {2025}),\ \Eprint {https://arxiv.org/abs/2509.10602} {arXiv:2509.10602 [hep-th]} \BibitemShut {NoStop}%
\bibitem [{\citenamefont {Stahl}\ \emph {et~al.}(2023)\citenamefont {Stahl}, \citenamefont {Qi}, \citenamefont {Glorioso}, \citenamefont {Lucas},\ and\ \citenamefont {Nandkishore}}]{Stahl:2023prt}%
  \BibitemOpen
  \bibfield  {author} {\bibinfo {author} {\bibfnamefont {C.}~\bibnamefont {Stahl}}, \bibinfo {author} {\bibfnamefont {M.}~\bibnamefont {Qi}}, \bibinfo {author} {\bibfnamefont {P.}~\bibnamefont {Glorioso}}, \bibinfo {author} {\bibfnamefont {A.}~\bibnamefont {Lucas}},\ and\ \bibinfo {author} {\bibfnamefont {R.}~\bibnamefont {Nandkishore}},\ }\href {https://doi.org/10.1103/PhysRevB.108.144509} {\bibfield  {journal} {\bibinfo  {journal} {Phys. Rev. B}\ }\textbf {\bibinfo {volume} {108}},\ \bibinfo {pages} {144509} (\bibinfo {year} {2023})},\ \Eprint {https://arxiv.org/abs/2303.09573} {arXiv:2303.09573 [cond-mat.stat-mech]} \BibitemShut {NoStop}%
\bibitem [{\citenamefont {Ganesan}\ and\ \citenamefont {Lucas}(2020)}]{Ganesan:2020wvm}%
  \BibitemOpen
  \bibfield  {author} {\bibinfo {author} {\bibfnamefont {K.}~\bibnamefont {Ganesan}}\ and\ \bibinfo {author} {\bibfnamefont {A.}~\bibnamefont {Lucas}},\ }\href {https://doi.org/10.1007/JHEP12(2020)149} {\bibfield  {journal} {\bibinfo  {journal} {JHEP}\ }\textbf {\bibinfo {volume} {12}},\ \bibinfo {pages} {149}},\ \Eprint {https://arxiv.org/abs/2008.09638} {arXiv:2008.09638 [hep-th]} \BibitemShut {NoStop}%
\bibitem [{\citenamefont {Nicolis}\ \emph {et~al.}(2014)\citenamefont {Nicolis}, \citenamefont {Penco},\ and\ \citenamefont {Rosen}}]{Nicolis:2013lma}%
  \BibitemOpen
  \bibfield  {author} {\bibinfo {author} {\bibfnamefont {A.}~\bibnamefont {Nicolis}}, \bibinfo {author} {\bibfnamefont {R.}~\bibnamefont {Penco}},\ and\ \bibinfo {author} {\bibfnamefont {R.~A.}\ \bibnamefont {Rosen}},\ }\href {https://doi.org/10.1103/PhysRevD.89.045002} {\bibfield  {journal} {\bibinfo  {journal} {Phys. Rev. D}\ }\textbf {\bibinfo {volume} {89}},\ \bibinfo {pages} {045002} (\bibinfo {year} {2014})},\ \Eprint {https://arxiv.org/abs/1307.0517} {arXiv:1307.0517 [hep-th]} \BibitemShut {NoStop}%
\bibitem [{\citenamefont {Nicolis}\ \emph {et~al.}(2015)\citenamefont {Nicolis}, \citenamefont {Penco}, \citenamefont {Piazza},\ and\ \citenamefont {Rattazzi}}]{Nicolis:2015sra}%
  \BibitemOpen
  \bibfield  {author} {\bibinfo {author} {\bibfnamefont {A.}~\bibnamefont {Nicolis}}, \bibinfo {author} {\bibfnamefont {R.}~\bibnamefont {Penco}}, \bibinfo {author} {\bibfnamefont {F.}~\bibnamefont {Piazza}},\ and\ \bibinfo {author} {\bibfnamefont {R.}~\bibnamefont {Rattazzi}},\ }\href {https://doi.org/10.1007/JHEP06(2015)155} {\bibfield  {journal} {\bibinfo  {journal} {JHEP}\ }\textbf {\bibinfo {volume} {06}},\ \bibinfo {pages} {155}},\ \Eprint {https://arxiv.org/abs/1501.03845} {arXiv:1501.03845 [hep-th]} \BibitemShut {NoStop}%
\bibitem [{\citenamefont {Esposito}\ \emph {et~al.}(2017)\citenamefont {Esposito}, \citenamefont {Garcia-Saenz}, \citenamefont {Nicolis},\ and\ \citenamefont {Penco}}]{Esposito:2017qpj}%
  \BibitemOpen
  \bibfield  {author} {\bibinfo {author} {\bibfnamefont {A.}~\bibnamefont {Esposito}}, \bibinfo {author} {\bibfnamefont {S.}~\bibnamefont {Garcia-Saenz}}, \bibinfo {author} {\bibfnamefont {A.}~\bibnamefont {Nicolis}},\ and\ \bibinfo {author} {\bibfnamefont {R.}~\bibnamefont {Penco}},\ }\href {https://doi.org/10.1007/JHEP12(2017)113} {\bibfield  {journal} {\bibinfo  {journal} {JHEP}\ }\textbf {\bibinfo {volume} {12}},\ \bibinfo {pages} {113}},\ \Eprint {https://arxiv.org/abs/1708.09391} {arXiv:1708.09391 [hep-th]} \BibitemShut {NoStop}%
\bibitem [{\citenamefont {Vegh}(2013)}]{Vegh:2013sk}%
  \BibitemOpen
  \bibfield  {author} {\bibinfo {author} {\bibfnamefont {D.}~\bibnamefont {Vegh}},\ }\href@noop {} {\bibinfo {title} {{Holography without translational symmetry}}} (\bibinfo {year} {2013}),\ \Eprint {https://arxiv.org/abs/1301.0537} {arXiv:1301.0537 [hep-th]} \BibitemShut {NoStop}%
\bibitem [{\citenamefont {Andrade}\ and\ \citenamefont {Withers}(2014)}]{Andrade:2013gsa}%
  \BibitemOpen
  \bibfield  {author} {\bibinfo {author} {\bibfnamefont {T.}~\bibnamefont {Andrade}}\ and\ \bibinfo {author} {\bibfnamefont {B.}~\bibnamefont {Withers}},\ }\href {https://doi.org/10.1007/JHEP05(2014)101} {\bibfield  {journal} {\bibinfo  {journal} {JHEP}\ }\textbf {\bibinfo {volume} {05}},\ \bibinfo {pages} {101}},\ \Eprint {https://arxiv.org/abs/1311.5157} {arXiv:1311.5157 [hep-th]} \BibitemShut {NoStop}%
\bibitem [{\citenamefont {Alberte}\ \emph {et~al.}(2018)\citenamefont {Alberte}, \citenamefont {Ammon}, \citenamefont {Jim\'enez-Alba}, \citenamefont {Baggioli},\ and\ \citenamefont {Pujol\`as}}]{Alberte:2017oqx}%
  \BibitemOpen
  \bibfield  {author} {\bibinfo {author} {\bibfnamefont {L.}~\bibnamefont {Alberte}}, \bibinfo {author} {\bibfnamefont {M.}~\bibnamefont {Ammon}}, \bibinfo {author} {\bibfnamefont {A.}~\bibnamefont {Jim\'enez-Alba}}, \bibinfo {author} {\bibfnamefont {M.}~\bibnamefont {Baggioli}},\ and\ \bibinfo {author} {\bibfnamefont {O.}~\bibnamefont {Pujol\`as}},\ }\href {https://doi.org/10.1103/PhysRevLett.120.171602} {\bibfield  {journal} {\bibinfo  {journal} {Phys. Rev. Lett.}\ }\textbf {\bibinfo {volume} {120}},\ \bibinfo {pages} {171602} (\bibinfo {year} {2018})},\ \Eprint {https://arxiv.org/abs/1711.03100} {arXiv:1711.03100 [hep-th]} \BibitemShut {NoStop}%
\bibitem [{\citenamefont {Baggioli}\ and\ \citenamefont {Grieninger}(2019)}]{Baggioli:2019abx}%
  \BibitemOpen
  \bibfield  {author} {\bibinfo {author} {\bibfnamefont {M.}~\bibnamefont {Baggioli}}\ and\ \bibinfo {author} {\bibfnamefont {S.}~\bibnamefont {Grieninger}},\ }\href {https://doi.org/10.1007/JHEP10(2019)235} {\bibfield  {journal} {\bibinfo  {journal} {JHEP}\ }\textbf {\bibinfo {volume} {10}},\ \bibinfo {pages} {235}},\ \Eprint {https://arxiv.org/abs/1905.09488} {arXiv:1905.09488 [hep-th]} \BibitemShut {NoStop}%
\bibitem [{\citenamefont {Baggioli}\ \emph {et~al.}(2020)\citenamefont {Baggioli}, \citenamefont {Grieninger},\ and\ \citenamefont {Li}}]{Baggioli:2020edn}%
  \BibitemOpen
  \bibfield  {author} {\bibinfo {author} {\bibfnamefont {M.}~\bibnamefont {Baggioli}}, \bibinfo {author} {\bibfnamefont {S.}~\bibnamefont {Grieninger}},\ and\ \bibinfo {author} {\bibfnamefont {L.}~\bibnamefont {Li}},\ }\href {https://doi.org/10.1007/JHEP09(2020)037} {\bibfield  {journal} {\bibinfo  {journal} {JHEP}\ }\textbf {\bibinfo {volume} {09}},\ \bibinfo {pages} {037}},\ \Eprint {https://arxiv.org/abs/2005.01725} {arXiv:2005.01725 [hep-th]} \BibitemShut {NoStop}%
\bibitem [{\citenamefont {Baggioli}\ and\ \citenamefont {Pujolas}(2015)}]{Baggioli:2014roa}%
  \BibitemOpen
  \bibfield  {author} {\bibinfo {author} {\bibfnamefont {M.}~\bibnamefont {Baggioli}}\ and\ \bibinfo {author} {\bibfnamefont {O.}~\bibnamefont {Pujolas}},\ }\href {https://doi.org/10.1103/PhysRevLett.114.251602} {\bibfield  {journal} {\bibinfo  {journal} {Phys. Rev. Lett.}\ }\textbf {\bibinfo {volume} {114}},\ \bibinfo {pages} {251602} (\bibinfo {year} {2015})},\ \Eprint {https://arxiv.org/abs/1411.1003} {arXiv:1411.1003 [hep-th]} \BibitemShut {NoStop}%
\bibitem [{\citenamefont {Armas}\ and\ \citenamefont {Have}(2024)}]{Armas:2023ouk}%
  \BibitemOpen
  \bibfield  {author} {\bibinfo {author} {\bibfnamefont {J.}~\bibnamefont {Armas}}\ and\ \bibinfo {author} {\bibfnamefont {E.}~\bibnamefont {Have}},\ }\href {https://doi.org/10.21468/SciPostPhys.16.1.039} {\bibfield  {journal} {\bibinfo  {journal} {SciPost Phys.}\ }\textbf {\bibinfo {volume} {16}},\ \bibinfo {pages} {039} (\bibinfo {year} {2024})},\ \Eprint {https://arxiv.org/abs/2304.09596} {arXiv:2304.09596 [hep-th]} \BibitemShut {NoStop}%
\bibitem [{\citenamefont {Yuan}\ \emph {et~al.}(2020)\citenamefont {Yuan}, \citenamefont {Chen},\ and\ \citenamefont {Ye}}]{Yuan:2019geh}%
  \BibitemOpen
  \bibfield  {author} {\bibinfo {author} {\bibfnamefont {J.-K.}\ \bibnamefont {Yuan}}, \bibinfo {author} {\bibfnamefont {S.~A.}\ \bibnamefont {Chen}},\ and\ \bibinfo {author} {\bibfnamefont {P.}~\bibnamefont {Ye}},\ }\href {https://doi.org/10.1103/PhysRevResearch.2.023267} {\bibfield  {journal} {\bibinfo  {journal} {Phys. Rev. Res.}\ }\textbf {\bibinfo {volume} {2}},\ \bibinfo {pages} {023267} (\bibinfo {year} {2020})},\ \Eprint {https://arxiv.org/abs/1911.02876} {arXiv:1911.02876 [cond-mat.str-el]} \BibitemShut {NoStop}%
\bibitem [{\citenamefont {G{\l}{\'o}dkowski}\ \emph {et~al.}(2024)\citenamefont {G{\l}{\'o}dkowski}, \citenamefont {Pe{\~n}a-Ben{\'\i}tez},\ and\ \citenamefont {Sur{\'o}wka}}]{Glodkowski:2024ova}%
  \BibitemOpen
  \bibfield  {author} {\bibinfo {author} {\bibfnamefont {A.}~\bibnamefont {G{\l}{\'o}dkowski}}, \bibinfo {author} {\bibfnamefont {F.}~\bibnamefont {Pe{\~n}a-Ben{\'\i}tez}},\ and\ \bibinfo {author} {\bibfnamefont {P.}~\bibnamefont {Sur{\'o}wka}},\ }\href {https://doi.org/10.1007/JHEP07(2024)285} {\bibfield  {journal} {\bibinfo  {journal} {JHEP}\ }\textbf {\bibinfo {volume} {07}},\ \bibinfo {pages} {285}},\ \Eprint {https://arxiv.org/abs/2401.01877} {arXiv:2401.01877 [hep-th]} \BibitemShut {NoStop}%
\end{thebibliography}%

\newpage
\appendix
\section{Equations of motion for linearized perturbations}\label{sectionA}

\subsection{Transverse sector}
Based on the Fourier decomposition defined as Eq.~\eqref{fourier} in the main text, the linear EOMs of the perturbations in the transverse sector are
\begin{align}
0=&-2 u^3 \bar{A}'_t a_{x}' -uh_{tx}'' +2  h_{tx}' +i k u  h_{xy}' -2 m^2 u \Phi _x'    \bar{V}',\\
0=&-u^4  h_{xy}  \bar{A}_t^{\prime 2}+u^2  h_{xy}  f'' +u^2 f'h_{xy}' -4 u  h_{xy}  f' +u^2 f   h_{xy}'' -2 u f   h_{xy}'\nonumber\\& +6 f   h_{xy} +i k u^2  h_{tx}' -2 i k u  h_{tx} +2 i u^2 \omega h_{xy}' -4 m^2 u^2h_{xy}    \bar{V}' +2 m^2h_{xy}    \bar{V}\nonumber\\& +2 \Lambda h_{xy} -2 i u \omega h_{xy} +2 i k m^2 u^2 \Phi _x    \bar{V}', \\
0=&-2 i  u^4 \omega  a_{x} \bar{A}'_t -  u^4  h_{tx} \bar{A}_t ^{\prime 2}+2 u  h_{tx}  f' -6 f   h_{tx} -i k u^2 f   h_{xy}' \nonumber\\&+2 m^2 u^2 f  \Phi _x'   \bar{V}' -i u^2 \omega   h_{tx}' +k^2 u^2  h_{tx} +2 m^2 u^2  h_{tx}    \bar{V}' -2 m^2  h_{tx}   \bar{V} \nonumber\\&-2 \Lambda h_{tx} +k u^2 \omega h_{xy} +2 i m^2 u^2 \omega  \Phi _x   \bar{V}'-u^4 \bar{A}'_t a^x_t ,\\
0=&\ m^2u f'\Phi _x'  \bar{V}' +2m^2 u^2 f  \Phi _x'   \bar{V}'' +m^2u f  \Phi _x''   \bar{V}' -2m^2 f  \Phi _x'   \bar{V}' +  m^2 u  h_{tx}'V' +2m^2 u^2  h_{tx}   \bar{V}'' \nonumber\\&-2m^2 h_{tx}   \bar{V}' -im^2    k u  h_{xy}  \bar{V}' -m^2k^2 u \Phi _x   \bar{V}' +2im^2 u^2 \omega  \Phi _x   \bar{V}'' +2im^2 u \omega  \Phi _x'V' \nonumber\\&-2im^2\omega  \Phi _x   \bar{V}'-u^3\bar{A}'_t a^{x\prime}_t,\\
0=&\  \bar{A}'_t   h_{tx}' +f  a_{x}'' + a_{x}' f' +2 i \omega a_{x}' -k^2  a_{x} +i  k  a^y_{x}-ika^x_{y}+a^{x\prime}_t,\\
0=&\ f a^{x\prime\prime}_t + a^{x\prime}_x  f' +2 i \omega a^{x\prime}_x -k^2  a^x_{x} ,\\
0=&-i k  a_{x} +f   a^{y\prime\prime}_{x} + a^{y\prime}_{x}  f' +2 i \omega a^{y\prime}_{x} -k^2 a^y_{x}-    a^y_{x}+a^x_{y},\\
0=&-a_x'-a^{x\prime\prime}_t +i k a^{x\prime}_y+\bar{A}'_t\Phi' _x,\\
0=&\ i k a^{x\prime}_t +f  a^{x\prime\prime}_y +a^{x\prime}_y  f' +2 i \omega a^{x\prime}_y- a^x_{y}+ika_x+a^y_{x} ,\\
0=&\ i \omega  a^{x\prime}_t -k^2 a^x_{t} - a^x_{t} +i k f  a^{x\prime}_y -k \omega a^x_{y}-i\omega a_x-\bar{A}'_t   h_{tx}-i\omega\bar{A}'_t \Phi_x-fa_x,
\end{align}
where the prime denotes the derivative of the radial coordinate $u$ for these perturbations, while it stands for the derivative of $X$ for $V$. We have adopted the radial gauge, i.e. $\delta g_{xu}=\delta A^x_u=0$.

\subsection{Longitudinal sector}
For the longitudinal sector, we introduce the symmetric combination $h_{xs}=(h_{xx}+h_{yy})/2$ and the anti-symmetric combination $h_{xa}=(h_{xx}-h_{yy})/2$ to simplify the equations. In the radial gauge, $\delta g_{uu}=\delta g_{tu}=\delta g_{yu}=\delta A_u=\delta A^y_u=0$, we have that
\begin{align}
0=&\ h_{xs}'' ,\label{eqtt}\\
0=&-2 u^4 f     \bar{A}_t'  a_t' -u^4 f  h_{xs} \bar{A}_t ^{\prime 2}+u^4 h_{tt} \bar{A}_t ^{\prime 2}-2 u h_{tt}  f' -i k u^2 h_{ty}  f' -u^2 f  f'  h_{xs}' \nonumber\\&-i u^2 \omega  h_{xs}  f' +2 u f  h_{xs}  f' -2 u f  h_{tt}' +12 f  h_{tt} +4 i k u f  h_{ty} +k^2 u^2 f  h_{xa} \nonumber\\&+4 u f ^2 h_{xs}' +k^2 u^2 f  h_{xs} +2 m^2 u^2 f  h_{xs}   \bar{V}'  -2  m^2 f  h_{xs}   \bar{V}  -2 \Lambda  f  h_{xs}\nonumber\\& +4 i u \omega  f  h_{xs} -6 f ^2 h_{xs} -2 i    k    m^2 u^2 f   \Phi_y    \bar{V}'  -k^2 u^2 h_{tt} +2  m^2 h_{tt} \bar{V}  \nonumber\\&+2 \Lambda  h_{tt} +2 i u \omega  h_{tt} -2 k u^2 \omega  h_{ty} -2 u^2 \omega ^2 h_{xs} ,\\
0=&\ 2 u^3    \bar{A}_t'  a_{y}' +u h_{ty}'' -2 h_{ty}' +i k u h_{xa}' +i k u h_{xs}' +2 m^2 u  \Phi_y ' \bar{V}'  ,\\
0=&-2 i k u^4 a_t \bar{A}_t' -2 i u^4 \omega  a_{y} \bar{A}_t' -2    u^4 a^y_{t} \bar{A}_t'-u^4 h_{ty}     \bar{A}_t ^{\prime 2}+2 u h_{ty}  f' -6 f  h_{ty}\nonumber\\& +i k u^2 f  h_{xa}' +i k u^2 f  h_{xs}' +2 m^2 u^2 f   \Phi_y ' \bar{V}'  -i k u^2 h_{tt}' +2 i k u h_{tt}\nonumber\\& -i u^2 \omega  h_{ty}' +2 m^2 u^2 h_{ty} \bar{V}'  -2 m^2 h_{ty} \bar{V}  -2 \Lambda  h_{ty} -k u^2 \omega  h_{xa} \nonumber\\&-k u^2 \omega  h_{xs} +2 i m^2 u^2 \omega \Phi_y  \bar{V}'  ,\\
0=&\ 2 u^4    \bar{A}_t'  a_t' -u^4 h_{xa}  \bar{A}_t ^{\prime 2}+u^2 h_{xa}  f'' +u^2 f'  h_{xa}' -4 u h_{xa}  f' -u^2 f'  h_{xs}'\nonumber\\& +u^2 f  h_{xa}'' -2 u f  h_{xa}' +6 f  h_{xa} -u^2 f  h_{xs}'' +2 u f  h_{xs}' +u^2 h_{tt}'' -4 u h_{tt}' +6 h_{tt}\nonumber\\& -2 i k u^2 h_{ty}' +4 i k u h_{ty} +2 i u^2 \omega  h_{xa}' -4   m^2 u^2 h_{xa} \bar{V}'  +2    m^2 h_{xa} \bar{V}  \nonumber\\&+2 \Lambda  h_{xa} -2 i u \omega  h_{xa} -2 i u^2 \omega  h_{xs}' -2 m^2 u^4 h_{xs} \bar{V}'' +2 i u \omega  h_{xs}\nonumber\\& -2 i k m^2 u^2  \Phi_y  \bar{V}'  +2 i k m^2 u^4  \Phi_y  \bar{V}'',\\
0=&\ 2 u^4    \bar{A}_t'  a_t' +u^4 h_{xa}     \bar{A}_t ^{\prime 2}-u^2 h_{xa}  f'' -u^2 f'  h_{xa}' +4 u h_{xa}  f' -u^2 f'  h_{xs}' -u^2 f  h_{xa}''\nonumber\\& +2 u f  h_{xa}' -6 f  h_{xa} -u^2 f  h_{xs}'' +2 u f  h_{xs}' +u^2 h_{tt}'' -4 u h_{tt}' +6 h_{tt} -2 i u^2 \omega  h_{xa}' \nonumber\\&+4 m^2 u^2 h_{xa} \bar{V}'  -2    m^2 h_{xa} \bar{V}  -2 \Lambda  h_{xa} +2 i u \omega  h_{xa} -2 i u^2 \omega  h_{xs}'\nonumber\\& -2 m^2 u^4 h_{xs} \bar{V}''  +2 i u \omega  h_{xs} +2 i  k  m^2 u^2  \Phi_y  \bar{V}'  +2 i k m^2 u^4  \Phi_y  \bar{V}''  ,\\
0=&\ 2 u^4 \bar{A}_t'  a_t' +u^4 h_{xs} \bar{A}_t ^{\prime 2}+u^2 f'  h_{xs}' -2 u h_{xs}  f' -4 u f  h_{xs}' +6 f  h_{xs} +2 u h_{tt}' -6 h_{tt} \nonumber\\&+i k u^2 h_{ty}' -4 i k u h_{ty} -k^2 u^2 h_{xa} +2 i u^2 \omega  h_{xs}' -k^2 u^2 h_{xs} -2 m^2 u^2 h_{xs} \bar{V}'\nonumber\\&+2 m^2 h_{xs} \bar{V}  +2 \Lambda  h_{xs} -4 i u \omega  h_{xs} +2 i k m^2 u^2  \Phi_y  \bar{V}' , \label{eqtu}\\
0=&\ -2 u^3 \bar{A}_t'  a^{y\prime}_{t} +2 m^2 u f' \Phi_y ' \bar{V}'  +4    m^2 u^2 f   \Phi_y ' \bar{V}'' +2 m^2 u f   \Phi_y '' \bar{V}' -4 m^2 f \Phi_y ' \bar{V}' \nonumber\\& +2  m^2 u h_{ty}' \bar{V}'  +4 m^2 u^2 h_{ty} \bar{V}''  -4  m^2 h_{ty} \bar{V}'  +2 i k m^2 u h_{xa} \bar{V}'  -2 i k m^2 u^3 h_{xs} \bar{V}'' \nonumber\\&-2 k^2 m^2 u  \Phi_y  \bar{V}' -2  k^2 m^2 u^3  \Phi_y  \bar{V}''  +4 i  m^2 u^2 \omega   \Phi_y  \bar{V}''+4 i m^2 u \omega \Phi_y'\bar{V}'\nonumber\\& -4 i m^2\omega \Phi_y\bar{V}' \label{eqphi} ,
\end{align}

\begin{align}
0=& -\bar{A}_t'  h_{xs}' -a_t'' +i k a_{y}' , \label{eqAt}  \\
0=&\ \bar{A}_t'  h_{ty}' +i ka_t' +f  a_{y}'' +a_{y}'  f' +2 i \omega  a_{y}' +   a^{y\prime}_{t}, \label{eqAy} \\
0=&\ i k h_{ty}  \bar{A}_t' +i \omega  h_{xs}  \bar{A}_t' +i \omega  a_t' -k^2 a_t +i k f  a_{y}' -k \omega a_{y} +i    k a^y_{t} , \label{eqAu} \\
0=&\ \bar{A}_t'   \Phi_y ' -   a_{y}' -a^{y\prime\prime}_{t} +i k a^{y\prime}_{y} , \label{eqAyt} \\
0=& -i k a^{y\prime}_{t} -f  a^{y\prime\prime}_{y} -a^{y\prime}_{y}  f' -2 i \omega  a^{y\prime}_{y}, \label{eqAyy} \\
0=&-   h_{ty}  \bar{A}_t' -i \omega   \Phi_y   \bar{A}_t' -i    k a_t -   f  a_{y}' -i \omega  a_{y} +i \omega  a^{y\prime}_{t} -k^2 a^y_{t}\nonumber\\&-   a^y_{t} +i k f  a^{y\prime}_{y} -k \omega  a^y_{y}. \label{eqAyu}
\end{align}

\section{Mode-decoupling and the subdiffusive collective mode at zero charge density}\label{sectionB}
\begin{figure}[htbp]
    \centering
\includegraphics[width=0.7
    \linewidth]{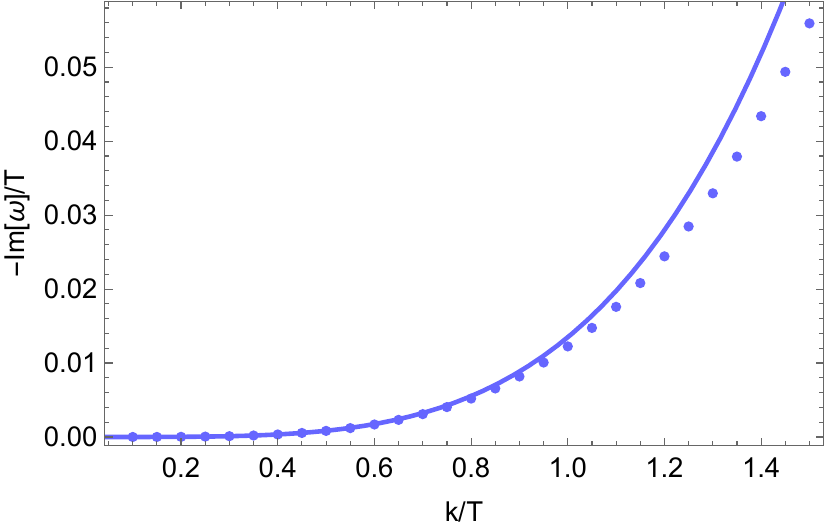}
    \caption{Dispersion relation of the subdiffusive mode given by the EOMs of $A$ and $B$, where the dots are QNMs data and the solid line represents the hydrodynamic fitting $\omega \sim -ik^4$. Here, we fix $\mu=0,\ m/T=0.2$.}
    \label{neutrallimit}
\end{figure}

In this appendix, we show that the appearance of the subdiffusion is unrelated to the metric and axion sectors. Considering the neutral case by turning off $\bar{A}_t$, it is found that the EOMs for the two gauge fields (Eq.~\eqref{eqAt}-Eq.~\eqref{eqAyu}) are decoupled from those of the metric (Eq.~\eqref{eqtt}-Eq.~\eqref{eqtu}) and the scalar field (Eq.~\eqref{eqphi}). We then yield that
\begin{align}
0=& -a_t'' +i k a_{y}' , \label{eqAt2}  \\
0=&\ ika_t' +f  a_{y}'' +a_{y}'  f' +2 i \omega  a_{y}' +   a^{y\prime}_{t}, \label{eqAy2} \\
0=&\ i \omega  a_t' -k^2 a_t +i k f  a_{y}' -k \omega a_{y} +i    k a^y_{t} , \label{eqAu2} \\
0=&-a_{y}' -a^{y\prime\prime}_{t} +i k a^{y\prime}_{y} , \label{eqAyt2} \\
0=&\ i k a^{y\prime}_{t}+f  a^{y\prime\prime}_{y} +a^{y\prime}_{y}  f' +2 i \omega  a^{y\prime}_{y}, \label{eqAyy2} \\
0=& -i    k a_t -   f  a_{y}' -i \omega  a_{y} +i \omega  a^{y\prime}_{t} -k^2 a^y_{t}-   a^y_{t} +i k f  a^{y\prime}_{y} -k \omega  a^y_{y}. \label{eqAyu2}
\end{align}
The EOMs above are not fully independent. One can easily check that Eq.~\eqref{eqAy2} and Eq.~\eqref{eqAyy2} can be derived from the other four equations and therefore are redundant. Taking the $u$-derivative of Eq.~\eqref{eqAu2} and Eq.~\eqref{eqAyu2} and combining them with Eq.~\eqref{eqAt2} and Eq.~\eqref{eqAyt2}, we obtain the two coupled equations,
\begin{align}
    0=&\ fA''+(2i\omega+f') A'-k^2A+ikB,\\
    0=&\ fB''+(2i\omega+f') B'-(k^2+1)B-ikA,
\end{align}
where the new variables $A \equiv a'_t$ and $B \equiv a^{y\prime}_t$ are introduced. The solution to these equations governs the subdiffusive mode shown in FIG.~\ref{neutrallimit}. 

\end{document}